\documentclass[superscriptaddress,
amsmath,amssymb,
aps,
pra,
floatfix,
nofootinbib,
twocolumn]{revtex4-2}

\usepackage{hyperref}
\usepackage{amsmath}
\usepackage{amssymb}
\hypersetup{colorlinks=true,
    linkcolor=blue,
    filecolor=blue, 
    citecolor=blue,
    urlcolor=blue,
}
\usepackage{graphicx}
\usepackage{braket}
\usepackage{dcolumn}
\usepackage{bm}
\usepackage{tabularx}
\usepackage{makecell}
\newcommand{\abs}[1]{\lvert #1 \rvert} 
\newcommand{\ev}[1]{\langle #1 \rangle} 
\newcommand{\lv}[0]{\mathcal{L}} 
\newcommand{\eps}[0]{\epsilon}

\begin{document}

\newcommand{\iitm}{\affiliation{Department of Physics, Indian Institute of Technology Madras, Chennai, India, 600036.}}
\newcommand{\itp}{\affiliation{Institute for Theoretical Physics, University of Innsbruck, 6020 Innsbruck, Austria}}
\newcommand{\iqoqi}{\affiliation{Institute for Quantum Optics and Quantum Information of the Austrian Academy of Sciences,  6020 Innsbruck, Austria}}
\newcommand{\iisc}{\affiliation{Department of Instrumentation and Applied Physics, Indian Institute of Science, Bangalore, India, 560012.}}
\newcommand{\fraunhofer}{\affiliation{Fraunhofer Institute for Applied Solid State Physics IAF, Tullastr. 72, 79108 Freiburg, Germany}}

\title{Exploring Quantum Synchronization with a Composite Two-Qubit Oscillator}

\author{Gaurav M. Vaidya}\email{gaurav.vaidya-1@colorado.edu}
\iitm
\author{Arvind Mamgain}\iisc
\author{Samarth Hawaldar}\iisc
\author{Walter Hahn}\itp\iqoqi\fraunhofer
\author{Raphael Kaubruegger}\itp\iqoqi
\author{Baladitya Suri}\iisc
\author{Athreya Shankar}\email{athreyas@iisc.ac.in}\itp\iqoqi\iisc
\date{\today}

\begin{abstract}
Synchronization has recently been explored deep in the quantum regime with elementary few-level quantum oscillators such as qudits and weakly pumped quantum Van der Pol oscillators. To engineer more complex quantum synchronizing systems, it is practically relevant to study \emph{composite} oscillators built up from basic quantum units that are commonly available and offer high controllability. Here, we consider a minimal model for a composite oscillator consisting of two interacting qubits coupled to separate baths, and show that this system exhibits a wide variety of synchronizing behaviors. We study the phase response of the constituent qubits as well as the system as a whole, when one of the qubits is weakly driven. We consider the thermal baths to have positive as well as effective negative temperatures, and discover effects that occur only when the temperatures of the baths for the two qubits are of opposite signs. We propose and analyze a circuit quantum electrodynamics implementation of this model, which exploits recent advances in dissipation engineering to realize effective negative temperature baths. Our work demonstrates the potential for assembling complex quantum synchronizing systems from basic building units, which is of pragmatic importance for advancing the field of quantum synchronization. 
\end{abstract}

\maketitle

\section{\label{sec:Introduction}Introduction}

Synchronization is at the heart of a variety of phenomena in nature and finds important practical applications, e.g. in the working of pacemakers and lasers \cite{pikovsky2001universal}. At its core, synchronization is the tendency of a self-powered, or \emph{self-sustained} oscillator (SSO) to lock to an external phase reference. 
Quantum synchronization explores how the synchronizing tendency of SSOs is affected by the strong quantum mechanical effects that arise when the oscillators are scaled down in size and energy~\cite{lee2013quantum,walter2014quantum,lee2014entanglement,sonar2018squeezing,roulet2018quantum,roulet2018synchronizing,koppenhofer2019PRA,tan2022quantum,lorch2017quantum,nigg2018observing,solanki2022symmetries,laskar2020observation,krithika2022observation,koppenhofer2020quantum,jaseem2020generalized,ameri2015PRA,shen2023quantum,buca2022SciPost,Tindall_2020}.  

A quantum SSO can be realized as the low occupation limit of a classical SSO or as a finite-dimensional system where only a few states are available even in principle. 
An example for the former is a weakly pumped Van der Pol oscillator~\cite{lee2013quantum,walter2014quantum,lee2014entanglement,sonar2018squeezing}. On the other hand, a qudit with gain and damping serves as a realization of a finite dimensional quantum SSO~\cite{roulet2018quantum,roulet2018synchronizing,koppenhofer2019PRA,tan2022quantum}. 

An important class of problems in quantum synchronization research is to understand the synchronization of model quantum SSOs to an external drive or the mutual synchronization of two (or more) SSOs when they are coupled.
Recent work has uncovered genuine quantum features in the response of quantum SSOs,  such as entanglement \cite{lee2014entanglement,roulet2018quantum}, and quantum interference effects that lead to synchronization blockade \cite{lorch2017quantum,nigg2018observing,solanki2022symmetries}. 
Furthermore, quantum synchronization is beginning to gain experimental relevance with the demonstration of elementary synchronizing systems in a vapor of Rb atoms \cite{laskar2020observation}, in nuclear spin systems \cite{krithika2022observation}, as well as by a digital simulation on a quantum computer \cite{koppenhofer2020quantum}. 

Going beyond paradigmatic systems, theoretical studies have discovered novel quantum synchronization phenomena in a variety of systems with structured energy levels and exotic gain and loss channels~\cite{tan2022quantum,murtadho2023PRL}. However, it remains unclear how such systems can be realized in practice. Therefore, complementary efforts are needed that explore how a variety of quantum synchronizing systems can be realized in the lab. With current technology, a promising and scalable approach to achieve this is to consider the assembly of a quantum many-body oscillator--- which we refer to as a \emph{composite} oscillator--- using elementary building blocks that are available in today's experiments, such as qubits. Such a `bottom-up' approach provides a path to assemble quantum synchronizing systems with desired features by tuning the properties of the constituent building blocks.


In this paper, we study a minimal model of a composite many-body quantum SSO made from two interacting qubits. The qubits are each coupled to separate thermal baths that provide local gain and damping to power the SSO. We study the response of this system to a weak synchronizing drive applied to one of the qubits. In addition to characterizing the synchronization of the composite system using global metrics, complementary insights can also be gained by studying local observables that capture the response of the individual constituents to the external drive. Besides being easier to measure, local observables also provide a window into how system parameters affect the internal working of the composite system and can reveal interesting features in the response of individual constituents that may not be apparent in global synchronization measures, as we will see below.
Accordingly, we first study how the phase response of the individual qubits, i.e. their tendency to develop a phase relative to the external drive, is modified by virtue of their mutual interaction. Remarkably, we find that under certain conditions, the interplay of gain, damping and interaction can cause the local phase response to completely vanish, despite the presence of the drive. Next, we explore the tendency of the composite oscillator, as a whole, to synchronize to the drive, using a recently introduced generalized measure of quantum synchronization~\cite{jaseem2020generalized}, which quantifies the overall buildup of coherence in the system because of the drive. We find that based on the choice of gain and damping rates, the qubit-qubit interaction can either strongly enhance or suppress the coherence buildup compared to that of a single qubit, thus leading to a diverse range of synchronizing behaviors that can be controlled by the system parameters. Finally, in line with our motivation to study experimentally feasible systems, we propose a realization of our model on a circuit quantum electrodynamics (QED) platform, using transmons and microwave resonators. We perform a detailed master equation simulation of our proposed implementation and show that it operates as a two-qubit oscillator under experimentally feasible parameters. Importantly, our implementation is scalable, and can be extended to realize many-body quantum SSOs made of more than two qubits.

Although a model of two interacting qubits with local thermal baths has been studied before in other contexts~\cite{scala2011EPJD,brask2015NJP,hofer2017NJP,cattaneo2019NJP}, our work demonstrates a hitherto unexplored aspect of this system, namely, its utility as a tunable testbed to explore quantum synchronization phenomena in composite quantum systems. Accordingly, our study is distinctive in the kind of metrics we consider, the parameter regimes we study, as well as the experimental scheme we propose to access these parameter regimes.  First, our interest is not in the intrinsic steady-state of the undriven two-qubit system \emph{per se}, but instead, we focus on the emergence of coherence in this system when a weak external drive is applied to one of the qubits. Consequently, we study metrics quantifying phase response and quantum synchronization, that are non-zero only when an external drive is applied. Second, in order to explore the full range of possible effects, we allow for the the local baths to have positive (damping dominated) as well as \emph{negative} (gain dominated) temperatures. Remarkably, we uncover effects that only arise when the two baths are inverted, i.e. when their temperatures have opposite signs.   Accordingly, a third distinguishing aspect of our work is the proposed implementation, which exploits recent demonstrations of dissipation engineering with transmons~\cite{sokolava2021singleatom,sokolova2023PRA} in order to realize effective negative temperature baths.

This paper is organized as follows. In Section~\ref{sec:The Model}, we introduce our two-qubit oscillator model and discuss the metrics we use to quantify the phase response of individual qubits and the synchronization of the composite system. We use these metrics to study the two-qubit oscillator as the system parameters are varied in Section~\ref{sec:Results}. In Section~\ref{sec:cqed}, we propose and simulate a circuit QED realization of the two-qubit oscillator. We conclude with a summary and outlook in Section~\ref{sec:conc}. Relevant additional details and extensions are provided in the form of Appendices. In particular, our model can be generalized to higher dimensional spins, which we illustrate with the example of a two-qutrit oscillator in Appendix~\ref{appsec:spin_1}.

\section{\label{sec:The Model}Model and Methods}

\begin{figure}[!tb]
    \centering
  \includegraphics[width=
  \columnwidth]{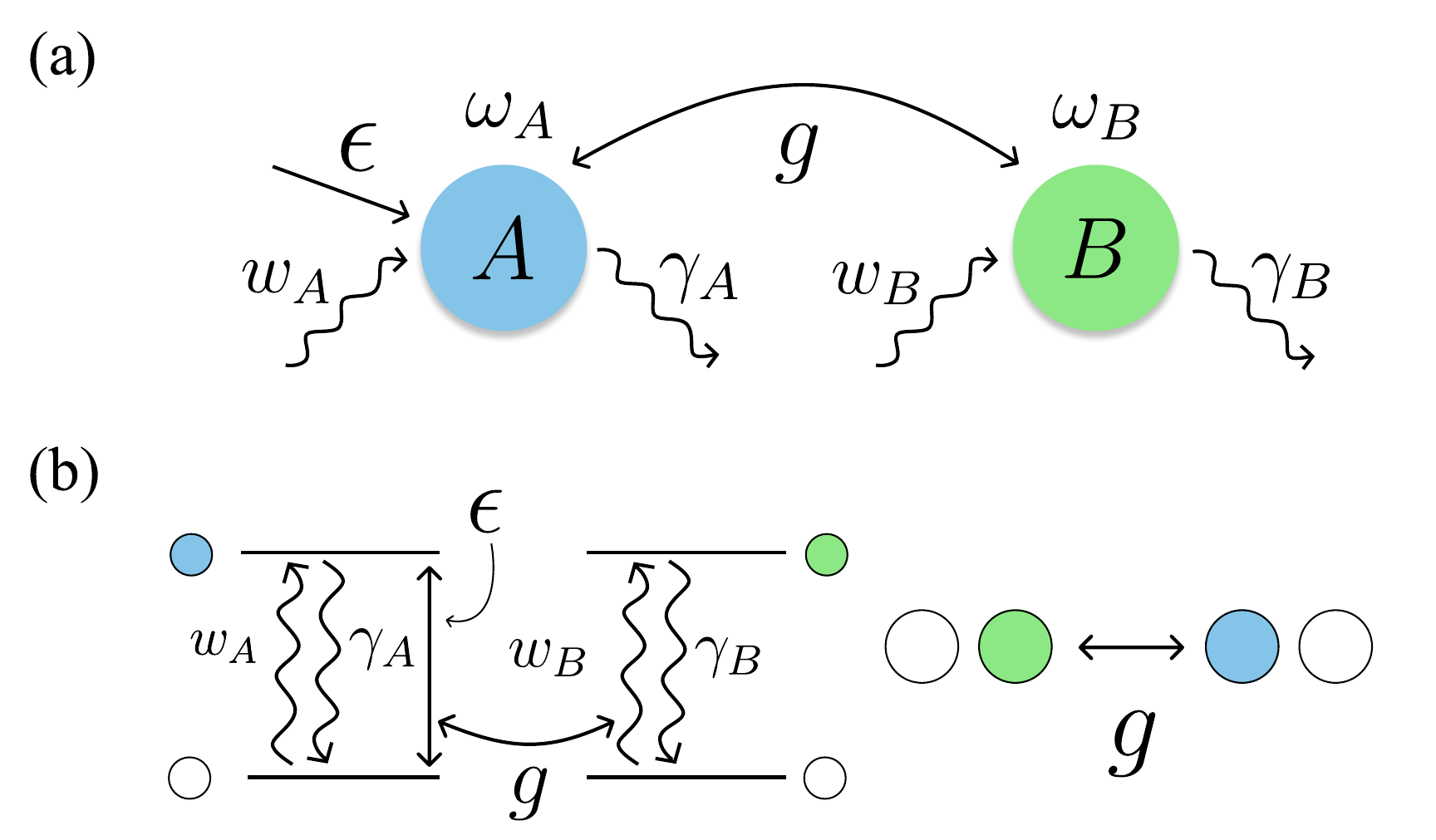}
    \caption{The composite two-qubit oscillator. (a) Schematic showing two interacting qubits, with each qubit coupled to a thermal bath. Additionally, qubit $A$ is weakly and coherently driven with strength $\epsilon$. (b) Energy level diagram showing the gain and damping channels, and the exchange interaction of the qubits. The right panel illustrates the flip-flop of excitations under the exchange interaction. The white (colored) circles represent the ground (excited) states of the two qubits.}
    \label{fig:model}
\end{figure}

In this section, we first describe the system under study. We then discuss the symmetry properties of the undriven steady state, and subsequently introduce the metrics we use to quantify the phase response of individual qubits and the synchronization of the composite oscillator.

\subsection{Model}

The model we consider is shown in Fig.~\ref{fig:model}. 
We consider two qubits $A$ and $B$, each with ground (excited) states $\ket{\downarrow}$ ($\ket{\uparrow}$) and with respective natural frequencies $\omega_A,\omega_B$, interacting via an exchange interaction with strength $g$. The Hamiltonian describing this system is~($\hbar=1$)
\begin{eqnarray}
    \label{eq:h_0}
    \hat{H}_0 &= &\omega_A\hat{S}_A^z+\omega_B\hat{S}_B^z + \frac{g}{2}\left(\hat{S}^{+}_A\hat{S}^{-}_B+\hat{S}^{-}_A\hat{S}^{+}_B\right). 
    \label{eqn:h_0}
\end{eqnarray}
Here $\hat{S}_j^z=(\ket{\uparrow}_j\bra{\uparrow}-\ket{\downarrow}_j\bra{\downarrow})/2$ for $j=A,B$ and $\hat{S}_j^\pm$ are raising and lowering operators defined as $\hat{S}_j^+=\ket{\uparrow}_j\bra{\downarrow}$ and  
$\hat{S}_j^-=\ket{\downarrow}_j\bra{\uparrow}$. Such a model can be realized on a number of platforms. For example, in the case of superconducting quantum circuits, a capacitive coupling between two transmon qubits gives rise to an interaction of the form $\hat{H}_\text{int}\propto g\hat{S}^x_A \hat{S}^x_B$, where $\hat{S}^x_j = (\hat{S}^+_j+\hat{S}^-_j)/2$. When $g,\abs{\Delta_q}\ll \omega_A,\omega_B$, where $\Delta_q=\omega_B-\omega_A$ is the relative qubit detuning, a rotating-wave approximation can be used to discard the $\hat{S}^+_A\hat{S}^+_B,\hat{S}^-_A\hat{S}^-_B$ terms to arrive at Eq.~(\ref{eqn:h_0}).

Furthermore, each qubit is weakly coupled to a thermal bath which leads to loss (gain) of excitations at rates $\gamma_j$ ($w_j$) with $j=A,B$. In particular, $\gamma_j>w_j$ ($\gamma_j<w_j$) corresponds to a positive (negative) temperature bath dominated by loss (gain). These two regimes occur on either side of an infinite temperature bath corresponding to $w_j=\gamma_j$. In order to describe the system using a Markovian master equation, the qubit-bath couplings have to be weak compared to $\omega_A,\omega_B$~\cite{cattaneo2019NJP}. Additionally, if the qubit-qubit coupling $g\ll\omega_A,\omega_B$ as considered here, the system can be accurately described using a \emph{local} master equation~\cite{cattaneo2019NJP,hofer2017NJP} given by 
\begin{eqnarray}
    \label{eq:master equation}
    \frac{d\hat{\rho}}{dt} &=& -i\left[\hat{H}_0, \hat{\rho}\right] \nonumber\\
    &&+ \sum_{j={A,B}}\mathfrak{D}[\sqrt{w_j}\hat{S}_j^+]\hat{\rho}+\sum_{j={A,B}}\mathfrak{D}[\sqrt{\gamma_j}\hat{S}_j^-]\hat{\rho},
\end{eqnarray}
where $\mathfrak{D}[\hat{O}]\hat{\rho} = \hat{O}\hat{\rho} \hat{O}^{\dagger}-\frac{1}{2}\hat{O}^{\dagger}\hat{O}\hat{\rho}-\frac{1}{2}\hat{\rho} \hat{O}^{\dagger}\hat{O}$ is the Lindblad dissipator. 

We note that previous studies have considered the \emph{mutual} synchronization of two quantum units when they are weakly coupled, i.e., when $g$ is weak compared to the gain and damping rates of the individual units~\cite{roulet2018quantum}. In contrast, here we consider the two-qubit system as a single composite oscillator whose intrinsic dynamics includes the qubit-qubit coupling, which is not restricted to be weak compared to the gain $w_A,w_B$ and damping rates $\gamma_A,\gamma_B$ of each qubit. The Lindblad master equation~(\ref{eq:master equation}) remains valid even when $g\gtrsim w_j,\gamma_j$, provided all of these parameters are very small compared to $\omega_A,\omega_B$~\cite{cattaneo2019NJP,hofer2017NJP}.

We study the response of this system to a weak synchronizing drive with frequency $\omega_d$ and strength $\epsilon$ that is applied to qubit $A$. When the drive detuning $\Delta_d=\omega_A-\omega_d \ll \omega_A$, the driving can be described under a rotating-wave approximation by the Hamiltonian
\begin{eqnarray}
    \label{eq:h_d}
    \hat{H}_d &= 
    \dfrac{\epsilon}{2} \left(\hat{S}^{+}_A e^{-i\omega_d t}+\hat{S}^{-}_A e^{i\omega_d t}\right). 
    \label{eqn:h_d}
\end{eqnarray}
To remove the time dependence, we transform the system into a frame rotating at $\omega_d$, where the total Hamiltonian is given by
\begin{eqnarray}
    \label{eq:h_tot}
    \hat{H}_\mathrm{tot} &= &\Delta_d\hat{S}_A^z+\left(\Delta_d+\Delta_q\right)\hat{S}_B^z\nonumber\\
    &&+\frac{g}{2}\left(\hat{S}^{+}_A\hat{S}^{-}_B+\hat{S}^{-}_A\hat{S}^{+}_B\right)
    +\frac{\epsilon}{2} \left(\hat{S}^{+}_A+\hat{S}^{-}_A\right).
    \label{eqn:h_tot}
\end{eqnarray}
The total master equation in the drive frame is given by Eq.~(\ref{eq:master equation}), with $\hat{H}_0$ replaced by $\hat{H}_\mathrm{tot}$. For discussing the results in Sec.~\ref{sec:Results}, we introduce characteristic frequencies 
\begin{eqnarray}
    \Gamma_j=\gamma_j+w_j, \; j=A,B,
    \label{eqn:Gamma}
\end{eqnarray} 
that define the total relaxation rate of a single qubit.

At a formal, mathematical level, the model described above can also apply to a two-\emph{qudit} oscillator wherein each spin has size $S>1/2$. In this general case, the levels of each spin can be labeled using the eigenstates $\ket{S,m}$ of the operator $\hat{S}^z$, which satisfy $\hat{S}^z\ket{S,m}=m\ket{S,m}$. The operators $\hat{S}^{\pm}$ are raising and lowering operators that transform the states according to $\hat{S}^\pm\ket{S,m}=\sqrt{(S\mp m)(S\pm m+1)}\ket{S,m\pm 1}$. In Appendix~\ref{appsec:spin_1}, we briefly study the case when $A$ and $B$ are qutrits, i.e. spins with $S=1$. We note, however, that 
the realization of such a system may be challenging and, furthermore, the specific form of the exchange interaction between qudits depends on the implementation. 

\subsection{Symmetry properties of the undriven steady state}
\label{sec:sym_undriven}

In order to understand the sense in which quantum synchronization occurs in the proposed model, it is useful to understand the symmetry properties of the master equation and the steady state of this system in the absence of the external drive.
Representing the master equation~(\ref{eq:master equation}) compactly as $\partial_t\hat{\rho} = \lv_0 \hat{\rho}$, where $\lv_0$ is the Liouvillian superoperator for the system, we observe that $\lv_0$ has a global $U(1)$ symmetry, i.e. it is invariant under the transformation $\lv_0\rightarrow \hat{U}(\phi)\lv_0 \hat{U}^\dag(\phi)$, where $\hat{U}(\phi)=e^{i\phi(\hat{S}^z_A+\hat{S}^z_B)}$. Consequently the steady state $\hat{\rho}_u$ satisfying $\lv_0\hat{\rho}_u=0$ also enjoys this symmetry, i.e. $\hat{U}(\phi)\hat{\rho}_u\hat{U}^\dag(\phi) = \hat{\rho}_u$. 

Writing the spectral decomposition $\hat{\rho}_u=\sum_{j=1}^4 \lambda_j \ket{\lambda_j}\bra{\lambda_j}$, the global $U(1)$ symmetry implies that each $\ket{\lambda_j}$ is an eigenstate of the operator $\hat{S}^z=\hat{S}^z_A+\hat{S}^z_B$. Accordingly, $\ket{\lambda_1}=\ket{\uparrow\uparrow}$ with eigenvalue $S^z=1$, $\ket{\lambda_4}=\ket{\downarrow\downarrow} (S^z=-1)$ and $\ket{\lambda_2},\ket{\lambda_3}$ are orthogonal linear combinations of $\ket{\uparrow\downarrow}$ and $\ket{\downarrow\uparrow}$ $(S^z=0)$. Their exact forms depend on the system parameters. Hence, $\hat{\rho}_u$ does not feature coherences between subspaces corresponding to different $S^z$ values, implying the absence of a preferred phase~\cite{jaseem2020quantum,krithika2022observation} between these subspaces. Synchronization in this system thus corresponds to the development of a preferred relative phase between these subspaces under the influence of an external, global $U(1)$ symmetry-breaking perturbation. Indeed, it is evident from the form of the drive in Eq.~(\ref{eqn:h_tot}) that, to leading order in $\eps$, it establishes coherences, i.e. phase relations, between subspaces with $\Delta S^z = \pm 1$.

\subsection{Phase response metric for individual qubits}
\label{sec:metric-individual}

In Sec.~\ref{sec: Phase Locking}, we study the phase response of the individual qubits constituting the system when a weak external drive is applied to one of them. The metric we use to quantify the phase response is the off-diagonal element, or coherence, of the steady-state reduced density matrix of each qubit. The choice of this metric is based on the phase space representation of the individual qubits using the Husimi $Q$ function, which we define here with respect to the $SO(3)$ coherent states~\footnote{We have omitted a normalization factor~\cite{roulet2018synchronizing} for simplicity. More generally, the $Q$ function for a spin-$S$ system can also be defined with respect to $SU(2S+1)$ coherent states~\cite{solanki2022symmetries,murtadho2023PRL,murtadho2023PRA}, but it is not required for our discussion here.}. For a general spin-$S$ system, the $Q$ function is defined as the overlap    
\begin{equation}
    \label{eq: Husimi}
    Q_{S,\hat{\rho}}(\theta, \varphi) = \bra{\theta,\varphi}\hat{\rho} \ket{\theta,\varphi}.
\end{equation}
Here, $\ket{\theta,\varphi}$ are the $SO(3)$ coherent states for a spin-$S$ system, which are defined via rotations of the state $\ket{S,m=S}$ as 
$\ket{\theta,\varphi} = e^{i\varphi  \hat{S}^z}e^{i\theta  \hat{S}^y}\ket{S,m=S}$~\cite{lee2015visualizing}.
The angles $\theta,\varphi$ respectively correspond to the polar and azimuthal angles on a generalized Bloch sphere. The $Q$ function therefore serves as a tool to visualize the state of the system on the surface of this sphere. 

For qubits ($S=1/2$), the $Q$ function can be expressed as 
\begin{eqnarray}
    Q_{\frac{1}{2},\hat{\rho}}(\theta,\varphi) &=& \frac{1}{2} + \ev{\hat{S}^z}\cos\theta +\mathrm{Re}[\ev{\hat{S}^+}e^{-i\varphi}]\sin\theta. \nonumber\\
    \label{eqn:q_half_spin}
\end{eqnarray}
The external drive introduces a nontrivial azimuthal phase distribution by establishing coherences in the system such that $\ev{\hat{S}^+}\neq 0$. This leads to a marginal distribution $P(\varphi)=(1/2\pi)\int_0^\pi d\theta\sin\theta \; Q_{\frac{1}{2},\hat{\rho}}(\theta,\varphi)$ for $\varphi$ that deviates from a uniform distribution. In Sec.~\ref{sec: Phase Locking}, we visualize this deviation by plotting the quantity $\delta P(\varphi)$ defined as 
\begin{eqnarray}
    \delta P(\varphi) = P(\varphi) - \frac{1}{2\pi} = \frac{1}{4}\mathrm{Re}[\ev{\hat{S}^+}e^{-i\varphi}].
    \label{eqn:delta_p}
\end{eqnarray}
Therefore, the phase response of individual qubits can be studied by probing the magnitude of the off-diagonal element of their reduced density matrices.

For a general spin-$S$ system, the $Q$ function can be decomposed into a sum of expectation values of spherical tensors, which is useful in studying the phase response of higher dimensional spin systems. We discuss this in more detail in Appendix~\ref{appendix:spherical_tensors}.

\subsection{Synchronization measure for composite oscillator}
\label{sec: metric-composite}

In Sec.~\ref{sec: Synchronisation}, we study the synchronization of the composite two-qubit oscillator, as a whole, when one of the qubits is weakly driven. For this study, we use an information theoretic measure of synchronization proposed in Ref.~\cite{jaseem2020generalized}. This metric is system agnostic, which makes it an attractive choice to study synchronisation of composite systems, where quasiprobability distributions may be inconvenient to compute as well as interpret. 

The central idea underlying this metric is to quantify synchronization as the deviation of the steady state $\hat{\rho}$ in the presence of the external drive from an appropriate limit cycle state $\hat{\rho}_\mathrm{lim}$ (described below), measured using a suitable measure of distance $\mathcal{D}$. In particular, when $\hat{\rho}_\mathrm{lim}$ is full rank, the distance $\mathcal{D}$ is taken as the relative entropy 
\begin{eqnarray}
    \mathcal{D}\equiv S(\hat{\rho} ||\hat{\rho}_\mathrm{lim}) = \mathrm{Tr}\left[\hat{\rho} \log(\hat{\rho})-\hat{\rho} \log(\hat{\rho}_\mathrm{lim})\right].
\end{eqnarray}

The limit cycle state $\hat{\rho}_\mathrm{lim}$ is the closest state to $\hat{\rho}$ which does not have the coherences induced by the drive. 
In general, the limit cycle state $\hat{\rho}_\mathrm{lim}$ is not just the steady state of the system when the drive is turned off. The reason is that, while the drive generally induces changes in populations as well as coherences, a synchronization metric must be sensitive only to the buildup of coherences. This subtlety is accounted for by minimizing $\mathcal{D}$ over an appropriate set, $\Sigma$, of `candidate' limit cycle states to obtain the synchronization measure, i.e. 
\begin{equation}
    \label{eq: optimise}
    \Omega(\hat{\rho}) = \min_{\hat{\rho}_\mathrm{lim} \in \Sigma} \mathcal{D}(\hat{\rho},\hat{\rho}_\mathrm{lim}),
\end{equation}
where the measure $\Omega(\hat{\rho})$ is called the relative entropy of synchronization. 

The set $\Sigma$ of candidate limit cycle states is chosen according to the system being studied. For our system, the global $U(1)$ symmetry discussed in Sec.~\ref{sec:sym_undriven} means that a natural choice of $\Sigma$ is the set of states that are diagonal in the eigenbasis $\{\ket{\lambda_j}\}$ of the steady state $\hat{\rho}_u$ of the \emph{undriven} oscillator, since synchronization to the external drive occurs via the buildup of coherences between the different bases $\{\ket{\lambda_j}\}$. For the particular case of such diagonal limit cycle states, the minimization in Eq.~(\ref{eq: optimise}) can be performed analytically and $\Omega(\hat{\rho})$ reduces to~\cite{jaseem2020generalized}
\begin{equation}
    \Omega(\hat{\rho}) = S(\hat{\rho}_\mathrm{diag})-S(\hat{\rho}).
    \label{eqn:sync-metric}
\end{equation}
Here, $S(\hat{\rho}) = \mathrm{Tr}\left[-\hat{\rho} \log (\hat{\rho})\right]$ is the von Neumann entropy and $\hat{\rho}_\mathrm{diag}$ is a state diagonal in $\{\ket{\lambda_j}\}$, obtained by simply deleting all the off-diagonal elements of $\hat{\rho}$ expressed in this basis.

We note that, for a system of the kind considered here,
Ref.~\cite{jaseem2020generalized} prescribes to choose $\Sigma$ as a set of so-called `partially-coherent' candidate limit cycle states. In Appendix~\ref{appendix:partially-coherent}, we show that such a choice leads to identical results as the ones we have obtained using diagonal limit cycle states, and provide an intuitive explanation for why this is the case.

\subsection{\label{sec:qualitative}Qualitative expectations}

\begin{figure}[!tb]
    \centering
  \includegraphics[width=
  0.8\columnwidth]{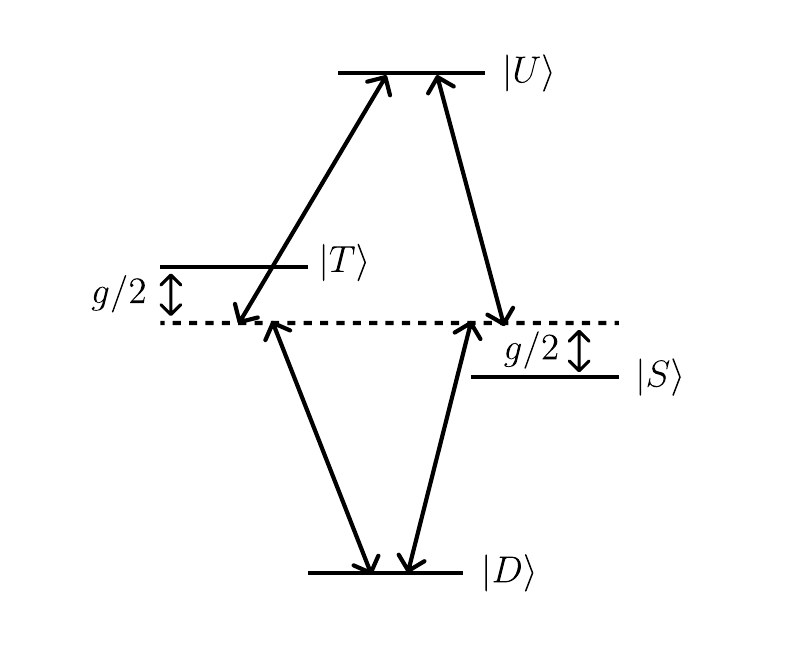}
    \caption{Effect of the weak drive in the dressed state representation.}
    \label{fig:model_singlettriplet}
\end{figure}

Before turning to the results, we provide some intuition for the behavior of the metrics introduced in the previous two subsections for different regimes of the qubit-qubit coupling. In the following qualitative discussion, we assume that $\Delta_d\approx\Delta_q\approx0$.

For $g/\Gamma \ll 1$ ($\Gamma_A\sim\Gamma_B\sim\Gamma$), the response of the oscillator to the drive is primarily just the response of qubit $A$, and the qubit-qubit coupling only leads to corrections of order $g/\Gamma$ or higher. On the other hand, when $g/\Gamma\gg 1$, the oscillator can be analyzed in the basis formed by the eigenstates of the undrived system Hamiltonian~\ref{eqn:h_0}. These are the usual dressed states, given by $\ket{D}=\ket{\downarrow\downarrow}$, $\ket{U}=\ket{\uparrow\uparrow}$, $\ket{T}=(\ket{\uparrow\downarrow}+\ket{\downarrow\uparrow})/\sqrt{2}$ and $\ket{S}=(\ket{\uparrow\downarrow}-\ket{\downarrow\uparrow})/\sqrt{2}$. The weak external drive is near resonance with the bare transition frequency of qubit $A$. In the dressed state picture, this drive translates to simultaneously driving four transitions as depicted by the black arrows in Fig.~\ref{fig:model_singlettriplet}. This can be seen by expressing the drive term in Eq.~(\ref{eqn:h_tot}) in the dressed basis, using the relation   
\begin{eqnarray}
    \hat{S}_A^+ = \frac{1}{\sqrt{2}}\left(\ket{U}\bra{T} - \ket{U}\bra{S} + \ket{T}\bra{D} + \ket{S}\bra{D} \right).
    \label{eqn:sigma_dressed}
\end{eqnarray}
For $g/\Gamma\gg 1$ ($\Gamma_A\sim\Gamma_B\sim\Gamma$), the states $\ket{T},\ket{S}$ are shifted from the bare resonance by $\pm g/2$ and hence all four transitions are driven off-resonantly. Hence, the coherence established in the system decreases and consequently the phase response and synchronization metrics asymptotically decay to zero as $g/\Gamma$ increases.

In the intermediate regime where $g/\Gamma\sim 1$, neither the individual qubit basis nor the dressed basis is particularly well suited for analyzing the system. This is because, while the Hamiltonian~(\ref{eqn:h_0}) is diagonal in the dressed basis, the jumps induced by the local baths are instead diagonal in the individual qubit basis. 
Hence, when the qubit-qubit coupling is comparable to the gain and loss rates, one can expect a complex interplay of these processes, that leads to non-trivial effects on the metrics. 

\section{\label{sec:Results} Results}

In Sec.~\ref{sec: Phase Locking}, we study the phase response of the individual qubits to the external drive and explore the role of the system parameters on their tendency to develop a phase relative to the drive. Subsequently, in Sec.~\ref{sec: Synchronisation}, we consider the two-qubit oscillator as a composite oscillator and study its collective response to the drive. An extension of this study to a two-qutrit oscillator is discussed in Appendix~\ref{appsec:spin_1}. In the following, we take the total relaxation rate [see Eq.~(\ref{eqn:Gamma})] of each qubit to be the same, i.e. $\Gamma_A=\Gamma_B=\Gamma$, and report frequency values ($w_j,\gamma_j,g,\Delta_d,\Delta_q,\epsilon, j=A,B$) in units of $\Gamma$, so that $\gamma_j = 1-w_j$ when expressed in these units.

\subsection{\label{sec: Phase Locking} Phase response of individual qubits}

\begin{figure}[!tb]
{\includegraphics[width=0.8\columnwidth]{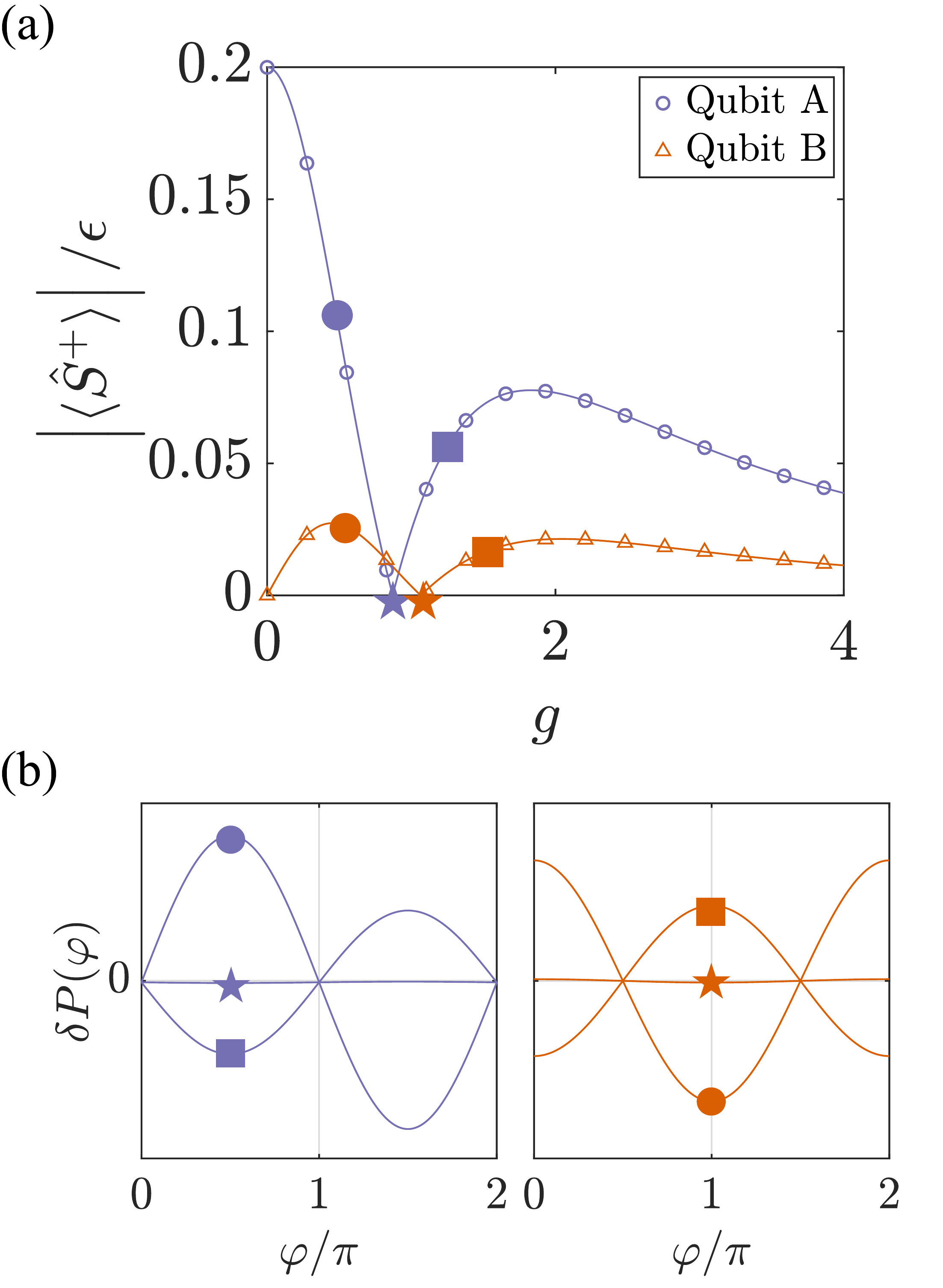}}%
\caption{Phase response of the individual qubits. (a) $\abs{\ev{\hat{S}^+}}/\epsilon$ for qubits A and B versus the coupling strength $g$ between the two qubits. Here, we choose $w_A = 0.4,\gamma_A=0.6$ and $w_B=0.75,\gamma_B=0.25$. Markers (lines) are numerical (analytical) results. (b) The variation in the azimuthal phase distribution $\delta P(\varphi)$ [see Eq.~(\ref{eqn:delta_p})] for qubit $A$ (left) and qubit $B$ (right) at the coupling strengths indicated in panel (a).}
\label{fig:vanishing-12}
\end{figure}

As described in Sec.~\ref{sec:metric-individual}, the phase response of a qubit is quantified by the magnitude of $\ev{\hat{S}^+}$, which is just the off-diagonal element, or coherence, of the reduced density matrix of the qubit. In Fig.~\ref{fig:vanishing-12}(a), we plot $\abs{\ev{\hat{S}^+}}$ normalized to the drive strength $\epsilon$ for both qubits as the qubit-qubit coupling strength $g$ is varied. Here, we have set $\Delta_d=\Delta_q=0$, i.e. the frequencies of the drive and the two qubits are taken to be equal. As $g$ increases, the phase response of qubit $A$ decreases and eventually vanishes completely at a particular strength $g_{0,A}$ indicated by the purple star. In the case of qubit $B$, we observe that it develops a non-zero phase response,  even though it is not directly driven, by virtue of its coupling with qubit $A$. Interestingly, the coherence of qubit $B$ also vanishes completely at a specific coupling strength $g_{0,B}$ (orange star). Finally, at large values of $g$, the coherence of either qubit approaches zero asymptotically, which can be understood as the result of off-resonant driving in the collective spin picture (Sec.~\ref{sec:qualitative}).

The complete vanishing of $\abs{\ev{\hat{S}_j^+}},j=A,B$ at $g_{0,j}$ corresponds to a zero \emph{crossing} of the quantity $\ev{\hat{S}_j^+}$, which in turn marks a $\pi$ phase shift in the phase developed by the corresponding qubit relative to the drive phase. We demonstrate this in  Fig.~\ref{fig:vanishing-12}(b), where we plot the variation in the azimuthal phase distribution $\delta P(\varphi)$ [see Eq.~(\ref{eqn:delta_p})] for the two qubits at coupling strengths before, at, and after the zero-crossing point. The distribution for either qubit is flat right at the zero crossing point while a $\pi$ phase shift is evident in the distributions before and after this point. 

The zero-crossing phenomenon occurs as a result of destructive interference from multiple drive pathways. For instance, in the case of qubit $A$, multiple pathways arise from the direct external driving and the feedback from qubit $B$ as a result of the coupling. Alternatively, one can also interpret this phenomenon as a destructive addition of coherences in the collective spin picture, as discussed in Sec.~\ref{sec:qualitative}. This phenomenon is intriguing because, for either qubit, the reduced density matrix at its respective zero-crossing point has an azimuthal phase symmetry as seen by the flat profile of $\delta P(\varphi)$, a remarkable feature given that the external drive explicitly breaks this symmetry in the system Hamiltonian~(\ref{eqn:h_tot}). Hence, in the following we will explore the parameter regimes where such a zero crossing can be observed.

\subsubsection{Zero crossing: Interplay of gain and loss rates}

The existence of a zero-crossing point is dependent on the temperatures of the local thermal baths coupled to each qubit. In order to rigorously determine the parameter regime where a zero-crossing can be observed, we first obtain an analytic expression for $\abs{\ev{\hat{S}_j^+}}, j=A,B$ as a function of the  coupling strength $g$, treating the drive strength $\epsilon$ as a perturbation. The details of this approach are presented in Appendix \ref{appendix:cumulant-approach}. As shown in Fig.~\ref{fig:vanishing-12}(a), the analytical expression (solid lines) is in excellent agreement with numerical results (markers). Next, we determine the existence of a zero-crossing point by solving for the  coupling strength $g_{0,j}$ where this expression vanishes. 

\begin{figure}[!tb]
{\includegraphics[width=0.8\columnwidth]{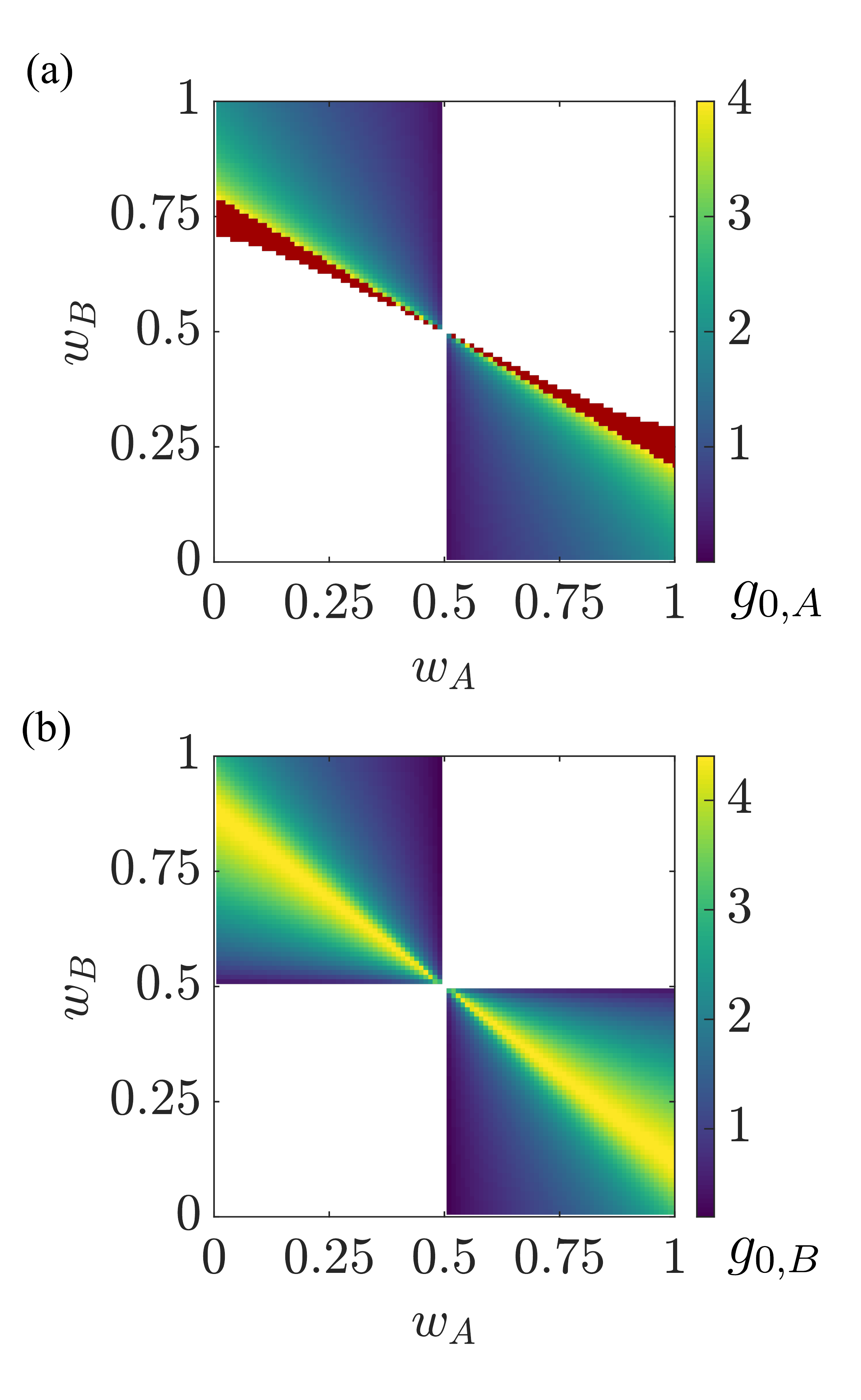}}
\caption{Bath parameters over which a zero crossing can be observed. The panels show the coupling strength at which the zero crossing occurs for (a) qubit $A$, and (b) qubit $B$, as $w_A,w_B$ are scanned while fixing $\gamma_A+w_A=\gamma_B+w_B=1$. The regions in white indicate the absence of a zero crossing, and the regions in red [in panel (a)] indicate values of $g_{0,A}>4$.}
\label{fig:parameter-sweep}
\end{figure}

In Fig.~\ref{fig:parameter-sweep}, we explore the existence of a zero-crossing point for each qubit as their gain rates (and consequently their bath temperatures) are varied. The color indicates the value of $g_0$, while the regions in white correspond to bath parameters where a zero-crossing point does not exist. We observe that, for both qubits, a zero-crossing point only exists when the baths are \emph{inverted} with respect to each other, i.e. when $w_A>\gamma_A$ and $w_B<\gamma_B$ or vice-versa. In other words, qubit $A$ (qubit $B$) must be coupled to a negative (positive) temperature bath or vice-versa. While this is a necessary condition to observe a zero crossing in qubit $A$, it is both necessary and sufficient in the case of qubit $B$. Furthermore, except in a narrow band (highlighted in red) for qubit $A$ where $g_{0,A}$ rapidly increases, the zero crossing typically occurs for values of $g_{0,j}\sim\mathcal{O}(1), j=A,B$, corresponding to the regime where qubit-qubit coupling strengths are comparable to the gain and loss rates of the qubits.

\subsubsection{Phase response to a detuned drive}

\begin{figure}[!tb]
{\includegraphics[width=0.8\columnwidth]{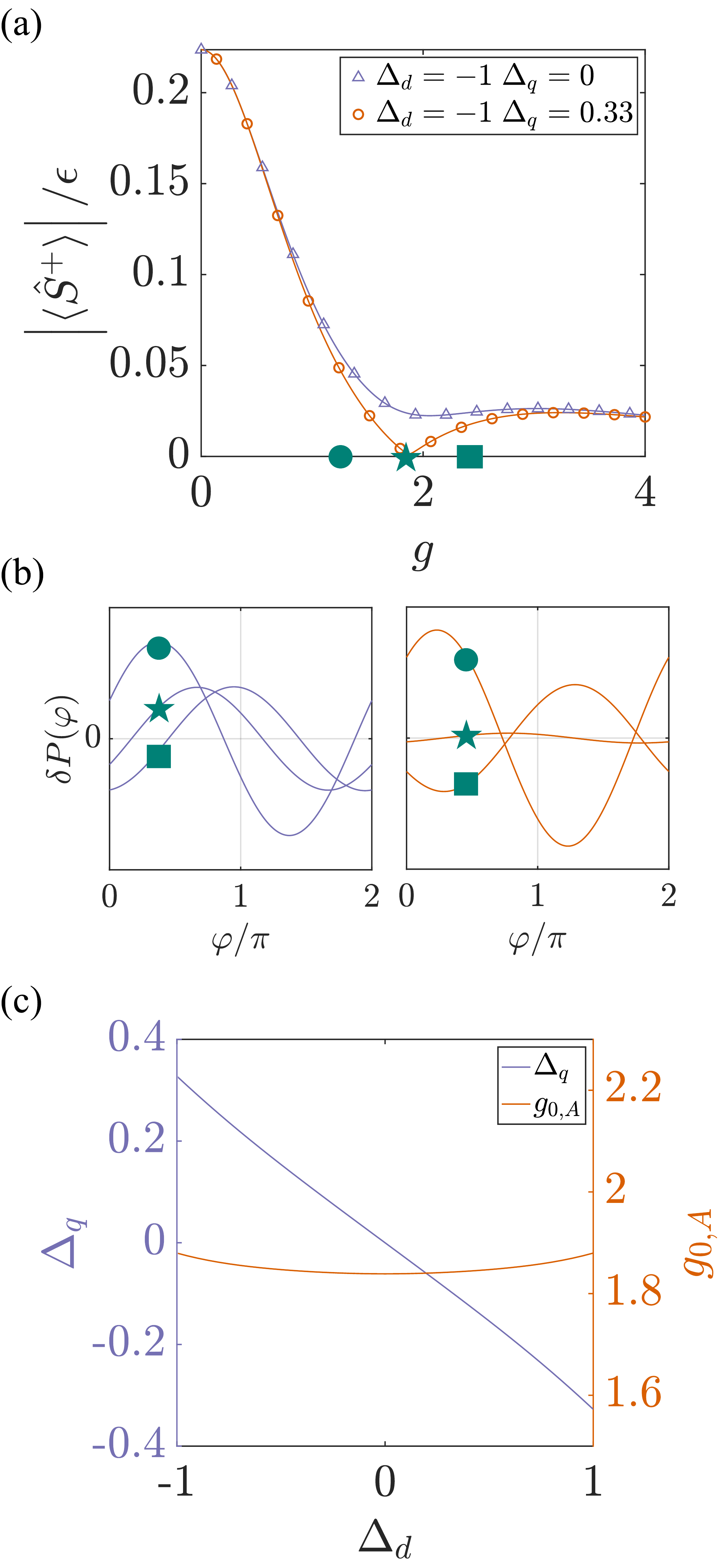}}
\caption{Phase response to a detuned drive. (a) $\abs{\ev{\hat{S}^+}}/\epsilon$ for qubit $A$ versus the coupling strength $g$ for a detuned drive. The purple (orange) curves correspond to the case when qubit $B$ is resonant with (detuned from) qubit $A$. Markers (lines) are numerical (analytical) results. (b) The variation in the azimuthal phase distribution $\delta P(\varphi)$ [see Eq.~(\ref{eqn:delta_p})] for qubit $A$ when qubit $B$ is resonant with (left) and detuned from (right) qubit $A$, at the coupling strengths indicated in panel (a). (c) The value of the qubit-qubit detuning $\Delta_q$ at which the zero crossing is restored, and the corresponding coupling strength $g_{0,A}$, as a function of the drive detuning $\Delta_d$. Here, we choose $w_A=0.25,\gamma_A=0.75$ and $w_B=0.75,\gamma_B=0.25$. 
}
\label{fig:parameter-sweep-delta}
\end{figure}

So far, we assumed that the drive is on resonance with qubit $A$. In Fig.~\ref{fig:parameter-sweep-delta}, we explore the phase response of qubit $A$ to the external drive when it is detuned. In Fig.~\ref{fig:parameter-sweep-delta}(a), we plot $\abs{\ev{\hat{S}_A^+}}$ for a detuned drive (purple curve) and find that the coherence no longer passes through a zero-crossing point (purple curve). To understand how this happens, we plot $\delta P(\varphi)$ at three different values of $g$ in the left panel of Fig.~\ref{fig:parameter-sweep-delta}(b). We observe that the location of the peaks and dips gradually shift to the right as $g$ increases, without ever passing through a flat profile. 

Interestingly, when qubit $B$ is appropriately detuned with respect to qubit $A$, i.e. $\Delta_q \neq 0$, we find that the zero-crossing point is restored, as seen in the orange curve in Fig.~\ref{fig:parameter-sweep-delta}(a). In the right panel of Fig.~\ref{fig:parameter-sweep-delta}(b), we plot $\delta P(\varphi)$ for $g$ values before, at, and after the zero-crossing point and find that, in contrast to the $\Delta_q=0$ case, the distribution passes through a flat profile similar to the case when $\Delta_q=\Delta_d=0$ [Fig.~\ref{fig:vanishing-12}(b)].

As shown in Fig.~\ref{fig:parameter-sweep-delta}(c), we find that for every drive detuning $\Delta_d$, there is a unique qubit-qubit detuning $\Delta_q$ that restores the zero-crossing point. Furthermore, the coupling strength $g_{0,A}$ at which this zero crossing occurs is essentially unchanged as the drive detuning is varied.  

These results provide a window into the internal dynamics of the composite two-qubit oscillator and demonstrate the role of system parameters such as bath temperatures, qubit-qubit interaction, and detunings in modifying the tendency of the constituent qubits to develop a preferred phase relative to the drive.

\subsection{\label{sec: Synchronisation} Synchronisation of the Composite System}

We now shift from the viewpoint of the individual qubits and instead study the response of the two-qubit system as a whole. We study the synchronization of the composite oscillator to the weak external drive applied to qubit $A$ by using the synchronization measure $\Omega(\hat{\rho})$ given in Eq.~(\ref{eqn:sync-metric}). In the following, we will drop the $\hat{\rho}$ dependency while referring to this measure for notational convenience.

\begin{figure}[!tb]
    \centering
    \includegraphics[width=0.80\columnwidth]{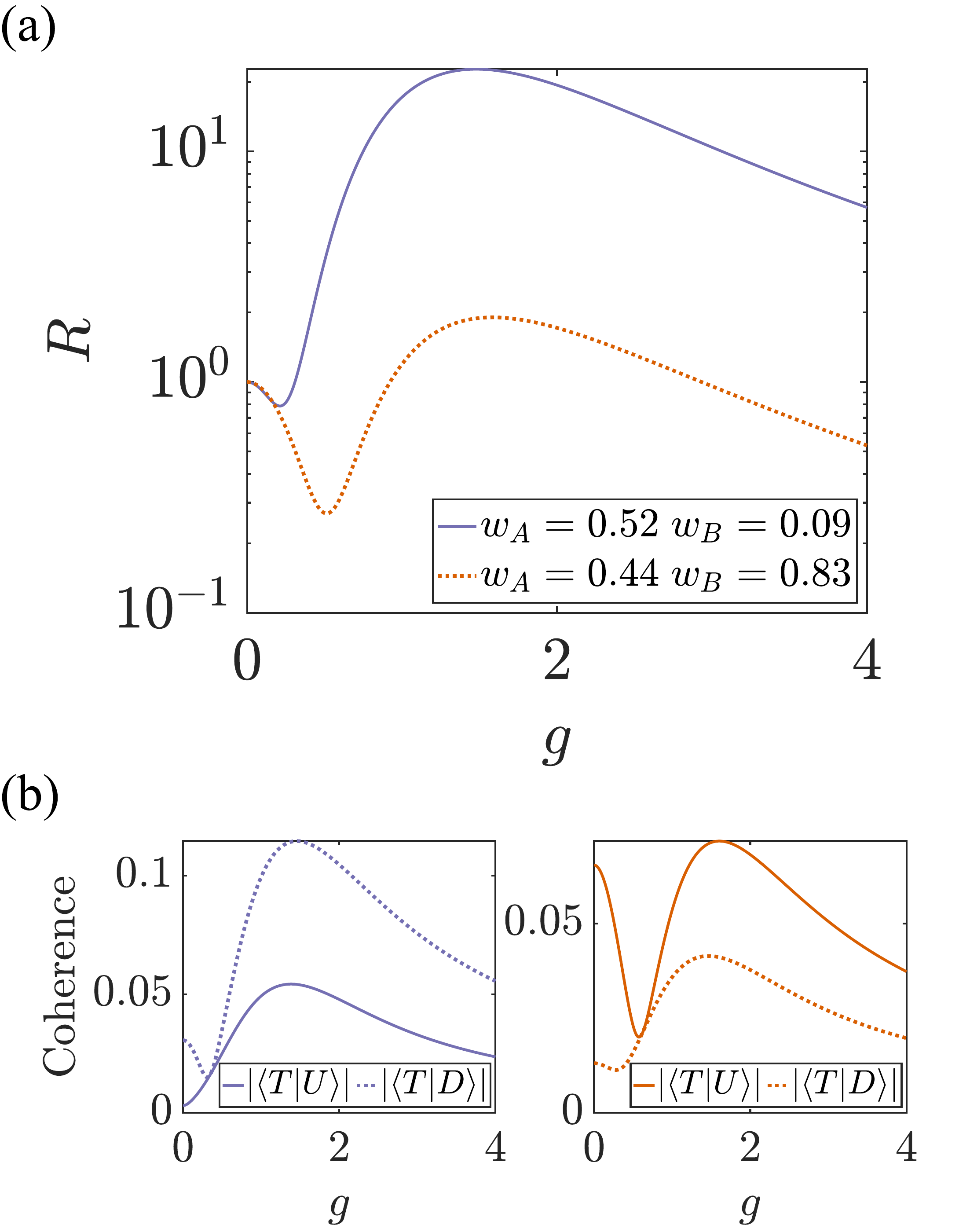}
    \caption{Enhancement and suppression of synchronisation of the composite two-qubit oscillator. The ratio $R$ [Eq.~(\ref{eqn:r_metric})] versus the coupling strength $g$ for two different sets of bath parameters, indicated by the $w_A,w_B$ values in the legend. The total relaxation rate for each qubit is kept fixed at $\gamma_A+w_A=\gamma_B+w_B=1$. (b) Magnitude of coherences in the collective spin basis (see Sec.~\ref{sec:qualitative}), $\abs{\ev{T|U}}$ and $\abs{\ev{T|D}}$, along the curves shown in panel (a).  Results are obtained numerically.}
    \label{fig:composite}
\end{figure}

We quantify the effect of qubit-qubit interactions on the synchronization to the drive via a ratio $R$, defined as the ratio of the values of $\Omega$ in the presence ($g\neq 0$) and absence ($g=0$) of qubit-qubit coupling, i.e.,  
\begin{eqnarray}
    R(g) = \frac{\Omega \rvert_{g}}{\Omega \rvert_{0}}.
    \label{eqn:r_metric}
\end{eqnarray}
In Fig.~\ref{fig:composite}, we plot $R$ versus $g$ for two different sets of bath parameters for the two qubits. The purple curve demonstrates that, for appropriate choices of gain and loss rates, qubit-qubit interactions can significantly enhance the extent of synchronization in a composite oscillator. On the other hand, interactions can also strongly suppress synchronization, as evidenced by the sharp dip in the orange curve.

The synchronization measure is sensitive to the steady-state coherences established by the drive in the composite oscillator. As a \emph{qualitative} indicator of this sensitivity, in Fig.~\ref{fig:composite}(b) we plot the quantities $\abs{\ev{T|U}}$ and $\abs{\ev{T|D}}$, corresponding to coherences in the dressed basis, along the curves displayed in Fig.~\ref{fig:composite}(a)~\footnote{We remind that the actual computation of the synchronization metric is done in the eigenbasis $\{\ket{\lambda_j}\}$ of the steady state of the undriven oscillator, as discussed in Sec.~\ref{sec: metric-composite}.}. The coherences $\ev{S|U}$ and $\ev{S|D}$ are respectively equal in magnitude to $\ev{T|U}$ and $\ev{T|D}$. We find that enhancement and suppression of the synchronization measure are qualitatively associated with corresponding peaks and dips in the magnitude of individual coherences in the dressed basis, demonstrating that the synchronization measure captures the overall extent of coherence build-up in the system as a result of the drive.

\subsubsection{Synchronization enhancement: Dependence on gain and loss rates}

\begin{figure}[!tb]
    \centering
    \includegraphics[width=0.80\columnwidth]{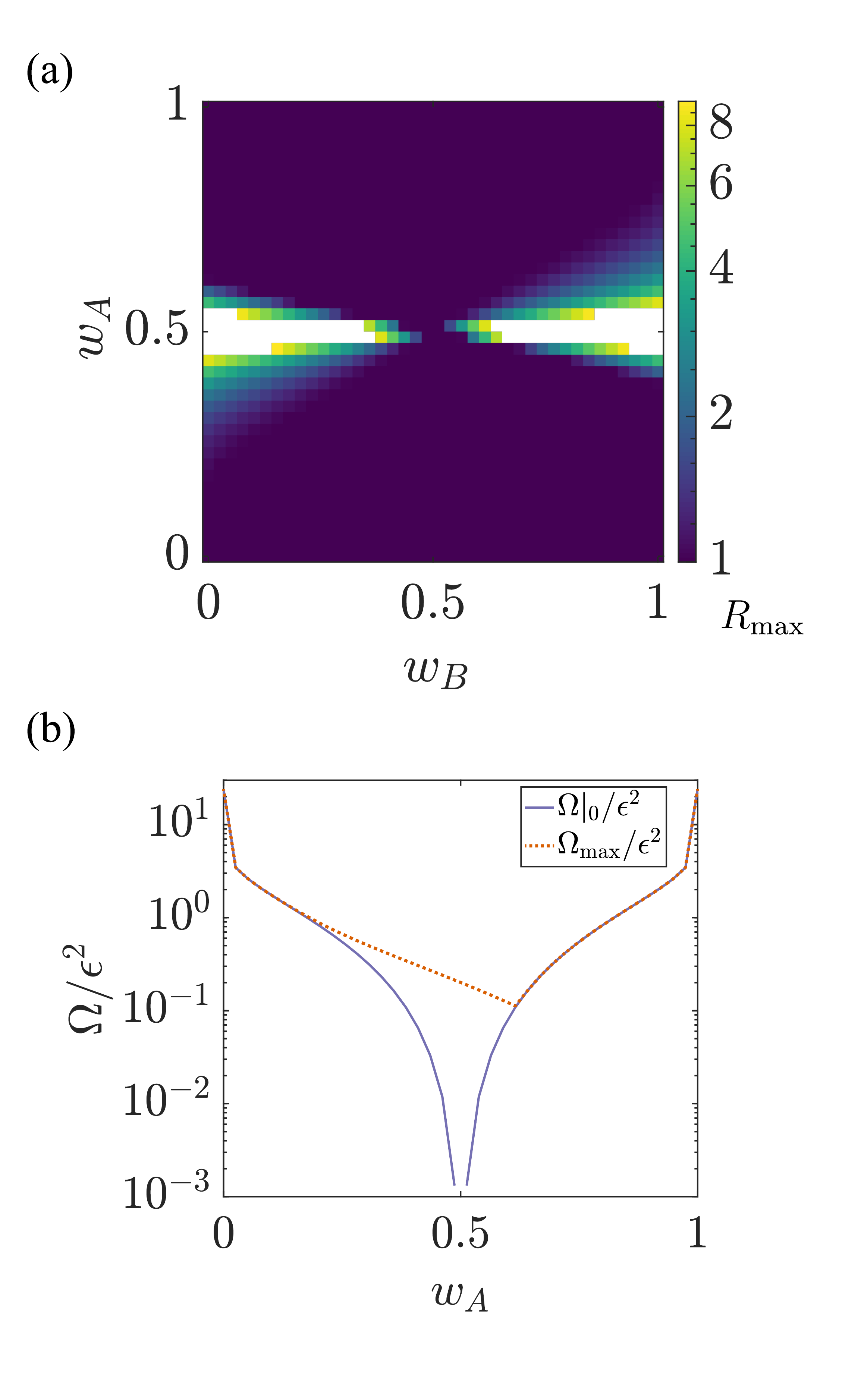}
    \caption{Effect of bath parameters on enhancement of synchronisation in the composite two-qubit oscillator. (a) Plot of $R_\mathrm{max}$, i.e. $R$ [Eq.~(\ref{eqn:r_metric})] maximized over coupling strength $g$, as $w_A,w_B$ are scanned. The regions in white correspond to $R_\mathrm{max}>10$. (b) Synchronization measure $\Omega$ [Eq.~(\ref{eqn:sync-metric})] versus $w_A$ for $w_B=0,\gamma_B=1$. The orange dashed (purple solid) line corresponds to $\Omega$ maximized over $g$ ($\Omega$ at $g=0$). In both panels, the total relaxation rate for each qubit is kept fixed at $\gamma_A+w_A=\gamma_B+w_B=1$. Results are obtained numerically.     
    }
    \label{fig:composite-parameter-sweep}
\end{figure}

The enhancement of synchronization as a result of qubit-qubit interaction depends on the parameters of the local thermal baths acting on each qubit. We once again first consider the situation when $\Delta_d=\Delta_q=0$. 
In Fig.~\ref{fig:composite-parameter-sweep}(a) we plot $R_\mathrm{max}$, the value of $R$ maximized over the coupling strength $g$, as the gain rates for the two qubits (and consequently their temperatures) are varied. We observe that significant enhancement in synchronization occurs when the gain and loss rates for qubit $A$ are comparable. This can be understood by considering the limiting case of $w_A=\gamma_A$, which corresponds to an infinite temperature bath. For $g=0$, the steady state of qubit $A$ coupled to such a bath is the maximally mixed state, which does not develop any coherence under an external drive. However, coupling it to a second qubit with $w_B\neq \gamma_B$ takes the composite oscillator away from infinite temperature and leads to a build up of non-zero coherence in the system.

To see the effect of the qubit-qubit coupling in the region $w_A \sim \gamma_A$ more clearly, we compare the case of coupled and uncoupled ($g=0$) qubits in Fig.~\ref{fig:composite-parameter-sweep}(b). We fix $w_B=0$ and plot $\Omega$ when it is maximized over $g$ ($\Omega_\mathrm{max}$), and for $g=0$ ($\Omega|_0$), as $w_A$ is varied. As $w_A\rightarrow\gamma_A$ ($w_A\rightarrow 0.5$ here), the coherence in the uncoupled system vanishes whereas it persists in the presence of interactions with qubit $B$. In fact, we find that for any non-zero temperature of qubit $A$, interactions with qubit $B$  with $w_B=0$ (zero temperature) lead to an enhancement in the synchronization measure, although this is most noticeable when $w_A\sim\gamma_A$.

We note that it is essential to keep in mind the actual value of the synchronization measure when interpreting enhancements in synchronization. For instance, as $w_A\rightarrow \gamma_A$, $R_\mathrm{max}\to\infty$. However, this result is an artefact of $\Omega|_0\to0$ in this limit, whereas $\Omega_\mathrm{max}$ remains finite, but small. Nevertheless, even a small non-zero buildup of coherences can lead to observable effects in macroscopic systems with a large number of quantum units, as occurs in NMR systems~\cite{krithika2022observation}.  

\subsubsection{Suppression of Synchronization}

\begin{figure}[!tb]
    \centering
    \includegraphics[width=0.80\columnwidth]{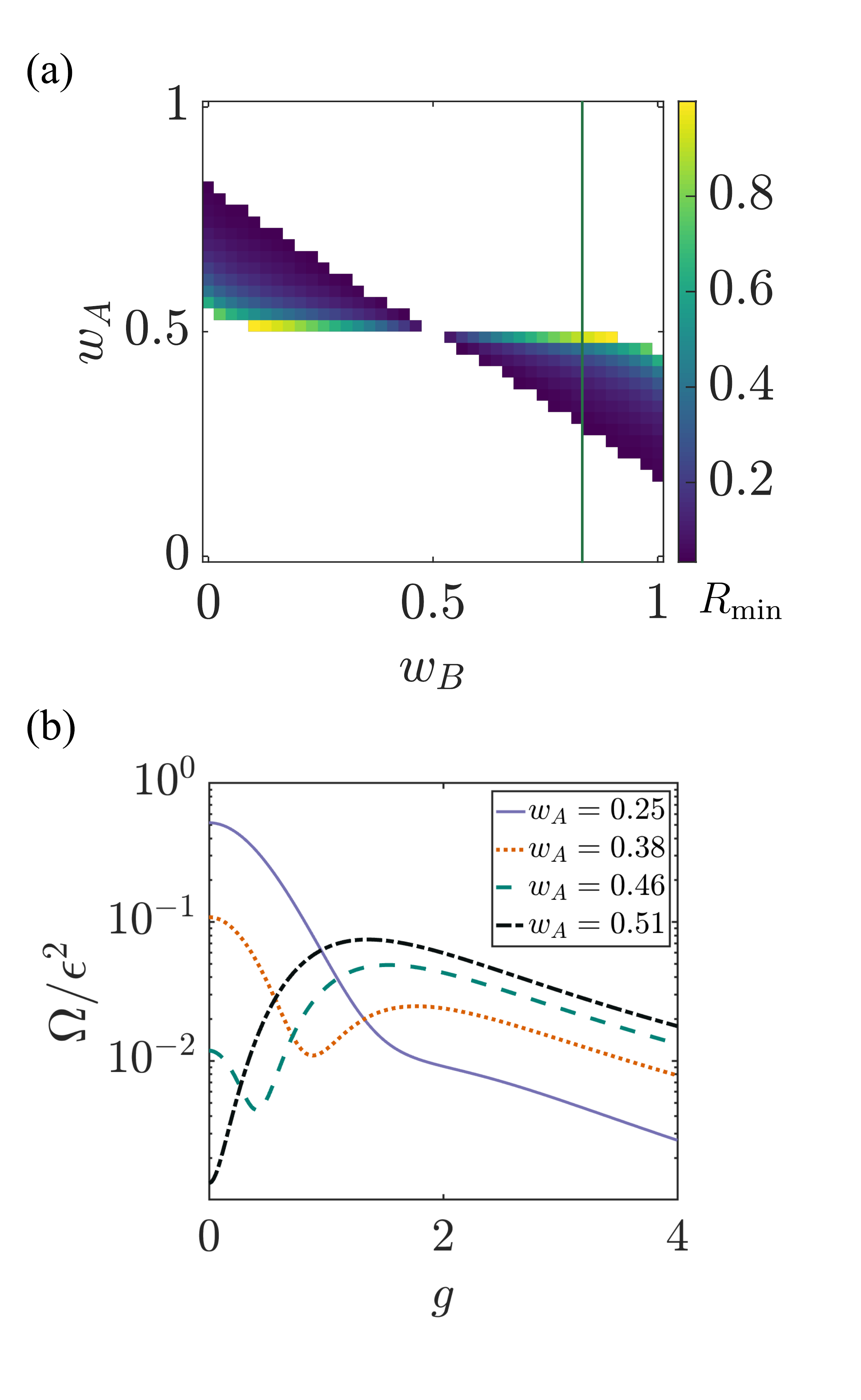}
    \caption{Effect of bath parameters on suppression of synchronization in the composite two-qubit oscillator. (a) Plot of $R_\mathrm{min}$, i.e. $R$ [Eq.~(\ref{eqn:r_metric})] evaluated at a local minimum as $g$ is scanned. White regions indicate absence of a local minimum. (b) Synchronization measure $\Omega$ [Eq.~(\ref{eqn:sync-metric})] versus $g$ evaluated at four different points along the vertical line shown in panel (a). In both panels, the total relaxation rate for each qubit is kept fixed at $\gamma_A+w_A=\gamma_B+w_B=1$. Results are obtained numerically.}
    \label{fig:supp}
\end{figure}

For $g\gg \Gamma$, $\Omega\to 0$ because of the large detuning of the weak drive, as discussed in Sec.~\ref{sec:qualitative}. However, Fig.~\ref{fig:composite}(a) shows that for appropriate bath parameters, a \emph{non-trivial} suppression of $\Omega$ can occur at intermediate values of $g$, which manifests as a local minimum of $\Omega$ (or equivalently $R$) as a function of $g$. In Fig.~\ref{fig:supp}(a), we explore the range of bath parameters over which such a local minimum exists by plotting the value of $R_\mathrm{min}$, i.e. the value of $R$ at the local minimum, whenever it exists. The white regions indicate the absence of a local minimum for those combinations of bath parameters. Interestingly, we find that a local minimum in synchronization occurs only in regimes where the baths for the two qubits are inverted with respect to each other, analogous to the existence of zero-crossing points in the phase response of the individual qubits. We note that the regimes of bath parameters for observing enhancement [Fig.~\ref{fig:composite-parameter-sweep}(a)] and suppression of synchronization [Fig.~\ref{fig:supp}(a)] are not mutually exclusive, because these effects occur at different values of the qubit-qubit coupling strength $g$, as can be seen from the orange line plot shown in Fig.~\ref{fig:composite}(a).

To understand how this minimum emerges and disappears, in Fig.~\ref{fig:supp}(b), we plot $\Omega/\eps^2$ versus $g$ for four points chosen along the vertical cut shown in Fig.~\ref{fig:supp}(a). For fixed $\omega_B\approx 0.82$, when $\omega_A$ is slightly above $0.5$, the $\Omega/\eps^2$ versus $g$ curve shows only a single maximum. As $\omega_A$ is decreased below $0.5$ to $\omega_A\approx0.46$, a local minimum can be observed. Further decreasing $\omega_A$ to $0.38$, the local minima and local maxima (at $g\approx 2$) become less pronounced. Finally, for $\omega_A\approx0.25$, $\Omega$ monotonically decays with $g$. 

These results demonstrate how the overall buildup of coherences in a composite system under external driving can be strongly enhanced or suppressed by tuning the parameters of the constituent quantum units and the interactions between them. More broadly, the variety of quantum synchronizing behaviors observable in our minimal model exemplifies the potential to assemble quantum self-sustained oscillators using basic building blocks such as qubits, which can then be used as a playground to explore aspects and applications of quantum synchronization~\cite{jaseem2020quantum}.

\subsection{Experimental considerations}

Motivated by practical considerations, we have studied the robustness of the features discussed above to qubit dephasing [see Eq.~(\ref{eq:dephase-master})] as well as stronger drive strengths such that $\epsilon/\Gamma\lesssim 1$. We find that both dephasing and stronger driving only lead to quantitative changes, e.g. in the location of the zero-crossing in Fig.~\ref{fig:vanishing-12}(a) or the extent of synchronization enhancement in Fig.~\ref{fig:composite}, but do not change the results qualitatively. 

\section{\label{sec:cqed}Proposal for Circuit QED Realization}

\begin{figure*}[!tb]
    \centering
    \includegraphics[width=0.6\linewidth]{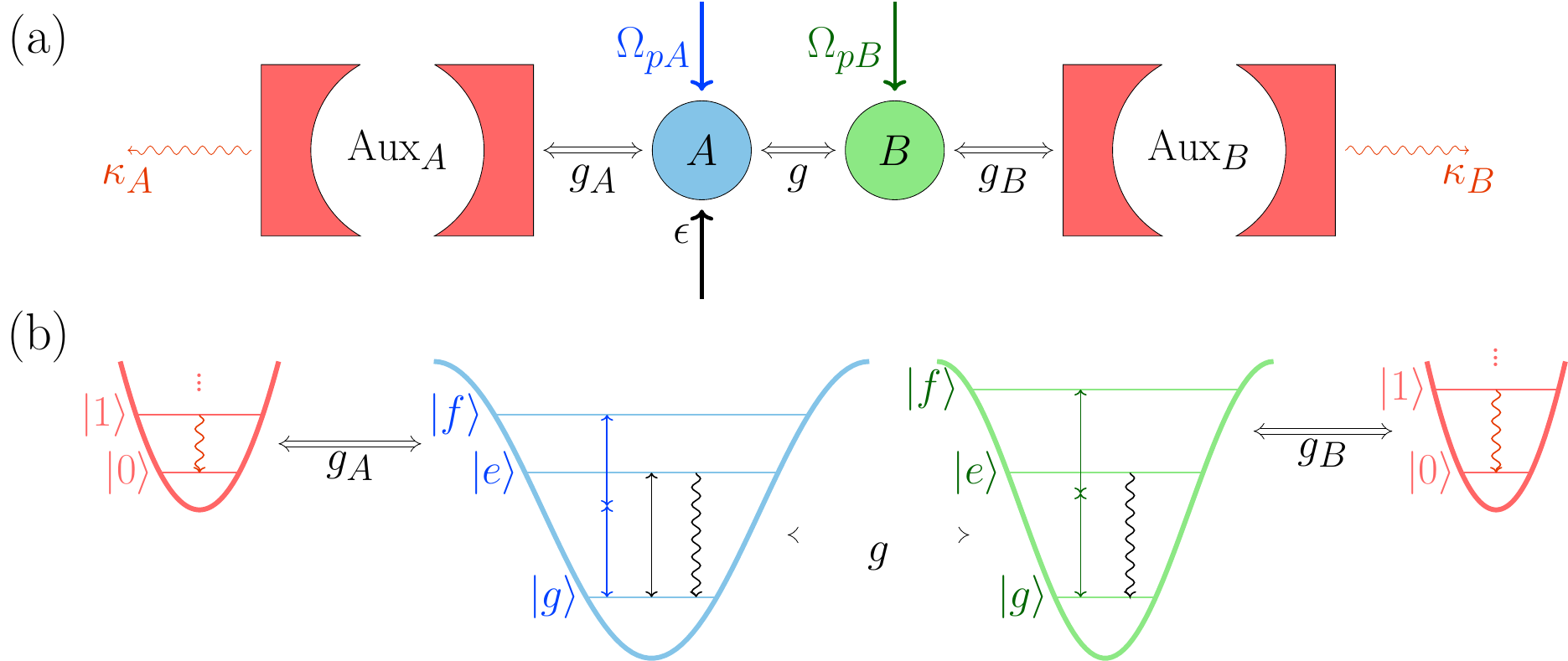}
    \caption{Circuit QED realization of a composite two-qubit oscillator. (a) Schematic showing two transmons $A$ and $B$ each coupled to a lossy auxiliary resonator as well as to each other, and driven using microwave fields. (b) Level diagram illustrating the working principle of the implementation.}
    \label{fig:cqed_setup}
\end{figure*}

In this section, we propose an implementation of our composite two-qubit oscillator model using artificial atoms and resonators made out of superconducting microwave circuits. This platform constitutes a favorable testbed to study synchronization for a number of reasons. These include the high degree of flexibility in the qubit connectivity, the ability to scale up the oscillator size if required, as well as the absence of certain undesirable effects such as motional heating, which often accompanies gain and loss channels in real atoms. In contrast to digital simulations of synchronizing systems on a superconducting quantum computer~\cite{koppenhofer2020quantum}, here we propose an analog simulation approach to directly engineer the various Hamiltonian and dissipative processes of the oscillator, as discussed below. 

Figure~\ref{fig:cqed_setup} shows a schematic of our proposed circuit QED implementation. The two qubits are encoded in the ground ($\ket{g}$) and first excited ($\ket{e}$) states of two tunable frequency transmons, labeled $A$ and $B$. While the loss channel, i.e. $\ket{e}\rightarrow \ket{g}$ decay, is intrinsic to each transmon, the gain channel, i.e. incoherent $\ket{g}\rightarrow \ket{e}$ jumps, need to be artificially engineered. Such a channel can be engineered by utilizing a lossy auxiliary resonator coupled to higher levels of the transmon~\cite{sokolava2021singleatom,sokolova2023PRA}. Specifically, for each transmon, a resonator with decay rate $\kappa_j$ ($j=A,B$) is coupled resonantly to its $\ket{e}\leftrightarrow\ket{f}$ transition. Exploiting the anharmonicity in the spacing of the transmon levels, two-photon $\ket{g}\rightarrow\ket{f}$ transitions can be driven resonantly by using appropriately detuned microwave `pump' fields $\Omega_{pA},\Omega_{pB}$. Consequently, population in $\ket{g}$ is transferred to $\ket{f}$, which decays rapidly to $\ket{e}$ as a result of coupling to the lossy resonator. The net effect of this process is an incoherent transfer of population from $\ket{g}$ to $\ket{e}$, that realizes a gain channel. The qubit-qubit interaction is realized using a tunable coupler (not shown) that introduces spin-exchange interactions with variable coupling strength $g$. Finally, the external drive $\epsilon$ is realized as an additional microwave field applied to transmon $A$.

To verify the realization of a two-qubit oscillator using this system, we have identified appropriate values for the various system parameters and performed master equation simulations of the circuit QED model. In our modeling, we include the Hilbert space of the transmons as well as the auxiliary resonators, while we choose to model the coupler as a phenomenological tunable coupling term between the two transmons. Our choice is motivated by the multiple demonstrations of tunable couplers \cite{li2020tunable,yan2018tunable,PhysRevX.11.021058}, making them a standard component in circuit QED systems. The details of the master equation simulations are discussed in Appendix \ref{appendix: details-sc} and the chosen parameter values are listed in Tables~\ref{tab:DeviceParams} and~\ref{tab:DeviceParams2}. These parameters are feasible with current technology.

The phase response and synchronization metrics can be measured in experiments. The off-diagonal matrix elements of the individual qubits can be straightforwardly measured by coupling individual readout resonators to each qubit  (not shown) and performing additional single qubit gates. Furthermore, the complete density matrix of the combined two-qubit system can be extracted in experiment by performing tomography using multiple single and two-qubit gates as done, e.g., in Ref.~\cite{li2017APL}. To infer the synchronization measure $\Omega$, the steady-state density matrices in the presence ($\hat{\rho}$) and absence ($\hat{\rho}_u$) of the external drive can be extracted, which together enable the construction of $\hat{\rho}_\text{diag}$. Subsequently, $\Omega$ [Eq.~(\ref{eqn:sync-metric})] can be evaluated by computing the entropies of the experimentally estimated  $\hat{\rho}$ and $\hat{\rho}_\text{diag}$ matrices.

\begin{figure}[!tb]
{\includegraphics[width=0.8\columnwidth]{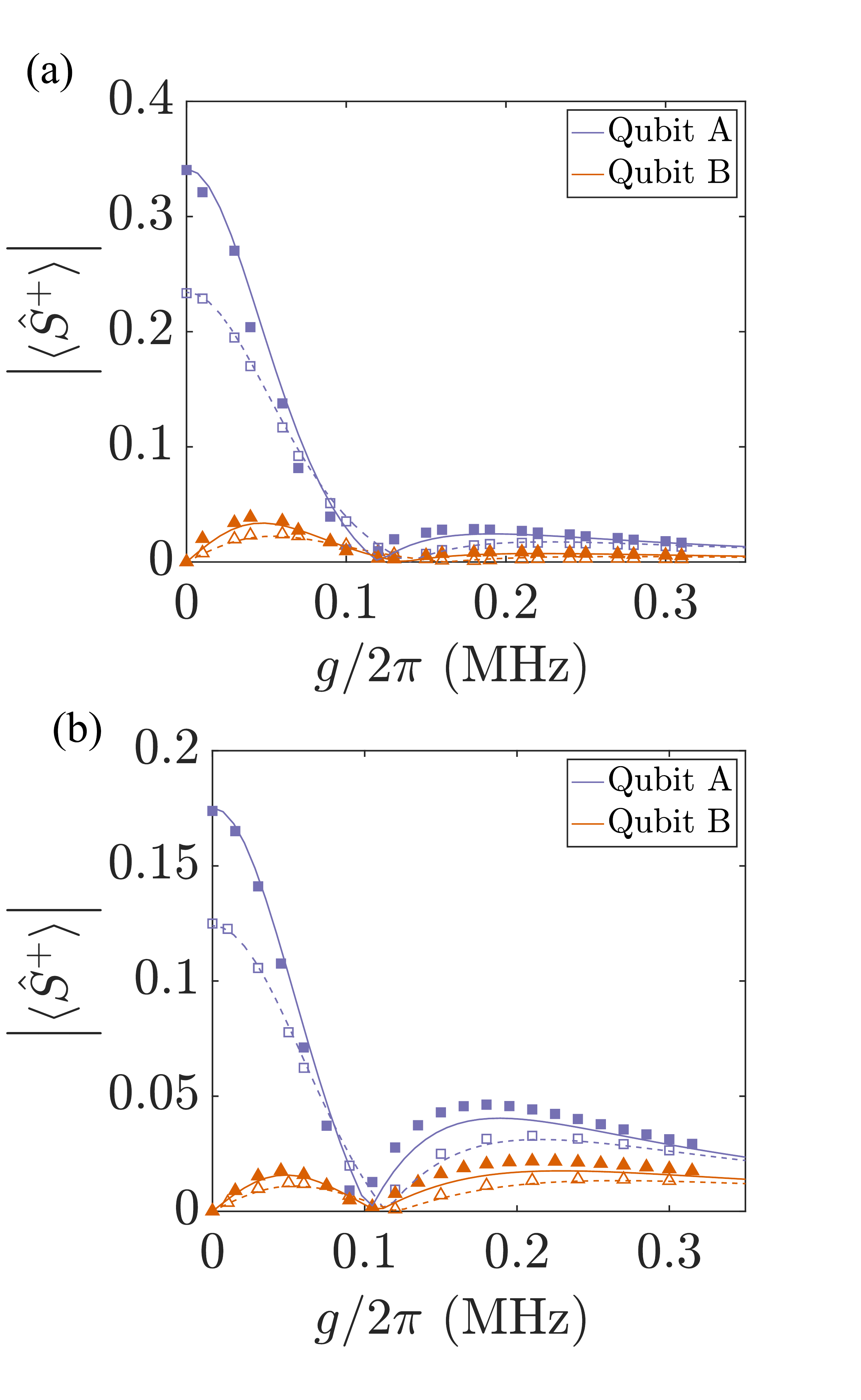}}
\caption{Phase response of individual qubits in the circuit QED implementation. The two panels show $\abs{\ev{\hat{S}^+}}$ for the two qubits with (a) $w_A/\gamma_A = 0$, $w_B/\gamma_B \approx 3.2$ and (b) $w_A/\gamma_A \approx 0.27$ and $w_B/\gamma_B \approx 4.5$. The filled markers (solid lines) show the results from the circuit QED model (qubit model, discussed in Sec.~\ref{sec:Results}). The empty markers (dashed lines) are the corresponding results in the presence of an additional dephasing channel. Parameters for the simulations are presented in Appendix~\ref{appendix: details-sc}.
}
\label{fig:sc-phasecorrelation}
\end{figure}

\begin{figure}[!tb]
{\includegraphics[width=0.8\columnwidth]{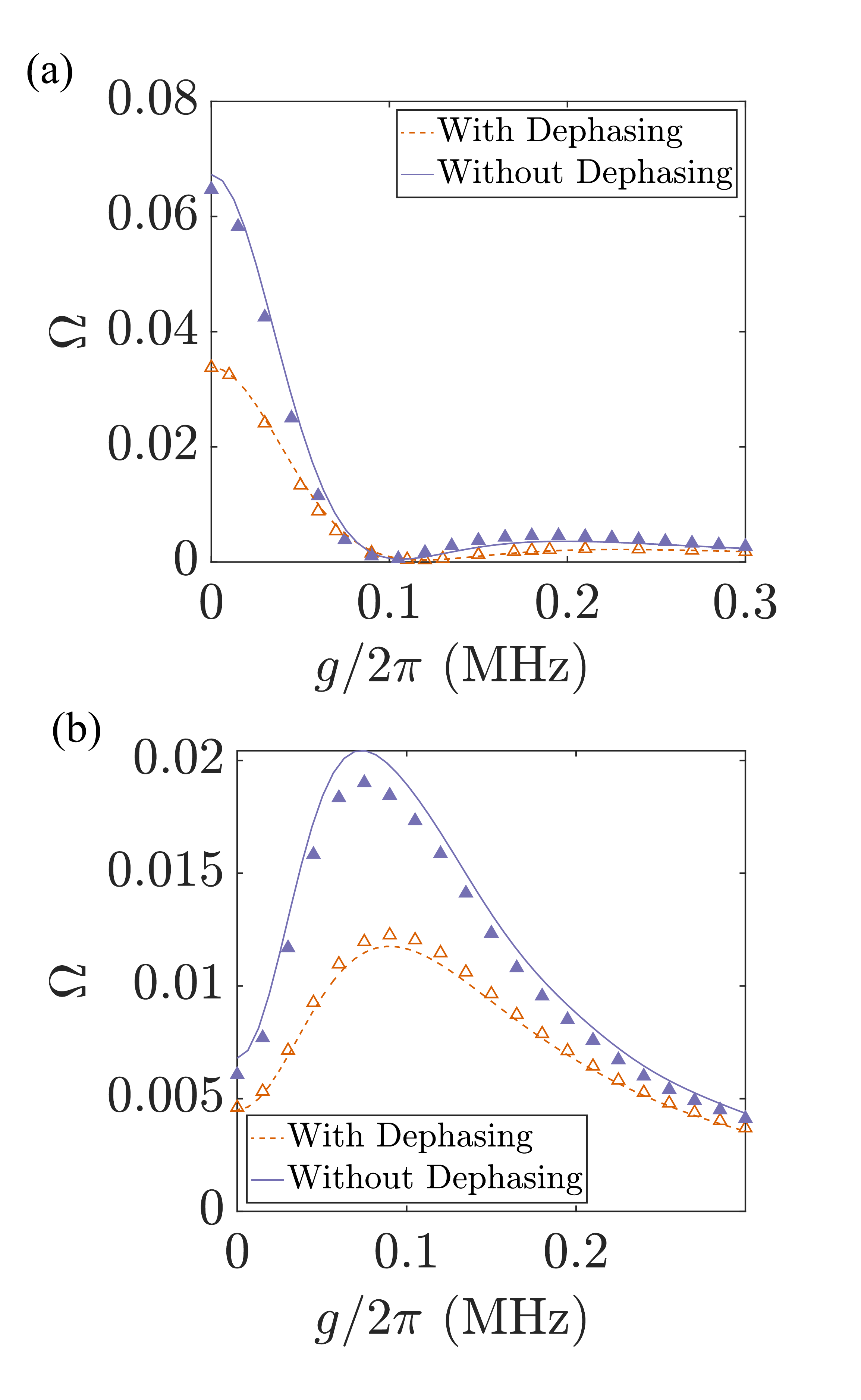}}
\caption{Synchronization of the composite oscillator in the circuit QED implementation. The two panels show the synchronization measure $\Omega$ [Eq.~(\ref{eqn:sync-metric})] for (a) $w_A/\gamma_A \approx 0.27$, $w_B/\gamma_B \approx 4.5$ and (b) $w_A/\gamma_A \approx 0.71$, $w_B/\gamma_B \approx 0.28$. The solid markers (solid lines) show the results for the circuit QED model (qubit model, discussed in Sec.~\ref{sec:Results}). The empty markers and dashed lines are the corresponding results in the presence of an additional dephasing channel. Parameters for the simulations are presented in Appendix~\ref{appendix: details-sc}.
}
\label{fig:sc-enhancement}
\end{figure}

We now turn to the simulation results, which demonstrate that our proposed system indeed operates as a composite two-qubit oscillator that can be used to explore the features discussed in Sec.~\ref{sec:Results}.
In Fig.~\ref{fig:sc-phasecorrelation}, we study the response of the individual qubits to the external drive by plotting the coherence $\abs{\ev{\hat{S}^+}}$ of the two qubits. The two panels correspond to two different sets of gain and loss rates for the two qubits. The results from the circuit QED model (markers) are in very good agreement with the expectations from the qubit model (lines) studied in Sec.~\ref{sec:Results}. In addition, we simulate the system in the presence of intrinsic transmon dephasing (empty markers and dashed lines) and find that it does not change the behavior qualitatively even when the dephasing rates are comparable to the relaxation rates $\gamma_A,\gamma_B$. In Fig.~\ref{fig:sc-enhancement}, we compare the synchronization measure $\Omega$ obtained from the circuit QED model to the predictions from the qubit model and once again find excellent agreement for two different sets of gain and loss parameters. Our results suggest that features such as the zero crossing in the coherence of the individual qubits and the enhancement or suppression of quantum synchronization as a result of qubit-qubit interactions can be observed in a circuit QED experiment, and are robust against effects such as dephasing. 

\section{\label{sec:conc}Conclusion and outlook}

We have introduced and studied a minimal model of a composite self-sustained oscillator consisting of two interacting qubits coupled to each other as well as to independent thermal baths. Such a model provides a first step towards engineering a wide variety of quantum synchronizing systems from basic units available on current quantum hardware. We studied the response of this system when a weak external drive is applied to one of the qubits. Specifically, we showed how the interplay of gain, loss and qubit-qubit interactions affects the phase response of the constituent qubits as well as the tendency of the composite system, as a whole, to synchronize to the drive. Furthermore, we demonstrated the experimental feasibility of our model by  proposing and analyzing a circuit QED implementation using transmons coupled to resonators as well as to each other.

Our study reveals that certain phase response and synchronization effects occur only when the baths for the two qubits are inverted, i.e. when gain dominates loss for one qubit and loss dominates gain for the other. In this situation, the phase response of the individual qubits to the external drive undergoes an abrupt phase shift of $\pi$  as the qubit-qubit coupling strength increases. Remarkably, at the crossover points for either qubit, which we term as zero-crossing points, we observe a blockade phenomenon: The off-diagonal element (coherence) of its reduced density matrix vanishes, restoring an azimuthal phase symmetry in the corresponding phase space distribution. We also studied the behavior of an information theoretic measure of quantum synchronization, which captures the tendency of the oscillator, as a whole, to synchronize to the drive. We find that when the gain and loss rates for the driven qubit are comparable, interactions with the second qubit can significantly enhance the coherence induced in the system by the drive. This enhancement occurs irrespective of whether the baths are inverted or not. On the other hand, a suppression of the synchronization response occurs only in parameter regimes where the baths are inverted, similar to the occurrence of zero-crossing points in the response of the constituent qubits of the system.

Our model naturally generalizes to higher dimensional spins, which may however be more challenging to implement in practice. In Appendix~\ref{appsec:spin_1}, we study a two-qutrit oscillator and show that the behavior of this system is qualitatively similar to the two-qubit oscillator. An interesting observation in the qutrit system is the occurrence of a zero crossing in the expectation values of higher-order spherical tensor operators at specific qutrit-qutrit 
coupling strengths, which can be interpreted as a generalized blockade phenomenon. 

Finally, let us note that our model and circuit QED proposal are complementary to, and build upon, previous studies of quantum synchronization with pairs of qubits in NMR platforms~\cite{krithika2022observation}. In contrast to these studies with an Ising interaction between the qubits, our model considers a spin-exchange type interaction between them. Furthermore, the qubit-qubit coupling strength and the individual qubit gain-to-loss ratios in our circuit QED proposal are tunable, allowing for control on the interactions and the local temperatures of each qubit. This enables the exploration of a wide variety of quantum synchronizing behaviors. More broadly, our proposal offers the potential to controllably scale up quantum self-sustained oscillators, and thereby experimentally probe the emergence of classical notions of synchronization from the underlying quantum system. For example, the properties of macroscopic synchronizing systems, such as superradiant lasers composed of several thousands to millions of  atoms~\cite{bohnet2012Nat,xu2014synchronization,weiner2017PRA} and analogous systems~\cite{zhu2015NJP,shankar2017PRA}, can be understood as the response of a large collective dipole, that can be analyzed with semiclassical mean-field type theories. Theoretical analysis of  scaled-up extensions of our model can be performed rigorously using recently introduced tools~\cite{landa2023SciPost}. Our proposal may also find applications in studying complex thermal heat engines~\cite{murtadho2023PRL,murtadho2023PRA} and exotic quantum heat engines, e.g., that operate between negative and positive temperature baths~\cite{bera2023arXiv}.

\section*{Acknowledgments}
We thank Sai Vinjanampathy and Simon Jäger for discussions and feedback on the manuscript. We thank 
 Peter Zoller for initial discussions on quantum synchronization. We acknowledge the use of QuTiP~\cite{johansson2012qutip} for numerical results. G. M. V. and S. H. acknowledge the support of Kishore Vaigyanik Protsahan Yojana, Department of Science and Technology, Government of India. A. M. acknowledges the support of Ministry of Education, Government of India. W. H. acknowledges support from a research fellowship from the DFG (Grant No. HA 8894/1-1). R. K. acknowledges funding from Large SCale Entangled Matter (LASCEM) via the US Air Force Office of Scientific Research grant no. FA9550-19-1-7044. B. S. acknowledges support of the Ministry of
Electronics and Information Technology, Government of
India, under the Centre for Excellence in Quantum Technology grant to Indian Institute of Science (IISc). A. S. acknowledges the support of a C. V. Raman Post-Doctoral Fellowship, IISc. 

\appendix

\section{\label{appendix:spherical_tensors} Phase response metrics based on the $Q$ function}

In Sec.~\ref{sec: Phase Locking}, we study the phase response of the individual qubits using the off-diagonal element of the respective reduced density matrices. This metric was motivated in Sec.~\ref{sec:metric-individual} using the Husimi $Q$ function. This approach can be generalized to study the phase response of individual qudits in a two-qudit oscillator by expanding the $Q$ function in terms of spherical tensor operators. Specifically, the $Q$ function can be expressed as the sum~\cite{dowling1994wigner,agarwal1981relation} 
\begin{equation}
    \label{eq: spherical}
    Q_S(\theta,\varphi) = \sum_{k=0}^{2S}\sum_{q=-k}^{k}c_{k,q}P_{k}^q(\cos \theta) e^{i q \varphi}\ev{\hat{T}_k^{-q}},
\end{equation}
where $P_k^q(\cos\theta)$ are associated Legendre polynomials,  $\hat{T}_k^q$ are spherical tensor operators, $\ev{\hat{O}} = \mathrm{Tr}\{\hat{O}\hat{\rho}\}$ denotes the expectation value of an operator $\hat{O}$, and $c_{k,q}$ are weight factors given by
\begin{equation}
    \label{eq: weights}
    c_{k,q} = \sqrt{\frac{(2k+1)(k-q) !}{(k+q) !}}\frac{(2S)!}{\left[(2S-k)!(2S+k+1)!\right]^{1/2}}.
\end{equation}
As a result, the phase response of individual qudits to an external drive that breaks the azimuthal phase symmetry can be studied by probing the expectation value of spherical tensor operators with $q$ different from zero.

In particular, for qubits  and qutrits, the $Q$ function can be explicitly expressed as 
\begin{widetext}
\begin{eqnarray}
    \label{equation:spin1and1/2Main}
    Q_{\frac{1}{2},\hat{\rho}}(\theta,\varphi) &=&  \frac{1}{\sqrt{2}}P_{0}^0 \ev{\hat{T}_0^0} + \frac{1}{\sqrt{2}}P_{1}^0 \ev{\hat{T}_1^0}
    +P_{1}^1 \mathrm{Re}\left[\ev{\hat{T}_1^1}e^{-i\varphi}\right], \nonumber\\ 
     Q_{1,\hat{\rho}}(\theta,\varphi) &=& \frac{1}{\sqrt{3}}P_{0}^0 \ev{\hat{T}_0^0} + \frac{1}{\sqrt{2}}P_{1}^0 \ev{\hat{T}_1^0}
    + \frac{1}{\sqrt{6}}P_{2}^0 \ev{\hat{T}_2^0} \nonumber\\
    &&+\mathrm{Re}\left[\left\{P_{1}^1 \ev{\hat{T}_1^1}
    +\frac{1}{3}P_{2}^1 \ev{\hat{T}_2^1}\right\}e^{-i\varphi} 
    +\frac{1}{6}P_{2}^2 \ev{\hat{T}_2^2}e^{-2i\varphi}\right],
\end{eqnarray}
\end{widetext}
where we have used $P_{k}^q \equiv P_k^q(\cos\theta)$ for compactness and expressed the $Q$ function using only the $q\geq 0$ terms. Writing the spherical tensor operators in $Q_{1/2}$ in terms of spin operators, we obtain Eq.~(\ref{eqn:q_half_spin}). In the case of spin-$1$, two multipoles $\hat{T}_1^{1}, \hat{T}_2^{1}$ contribute to the first harmonic in $\varphi$, while $\hat{T}_2^2$ gives rise to a second harmonic. We study these quantities in the context of a two-qutrit oscillator in Appendix~\ref{appsec:spin_1}.

\section{\label{appendix:partially-coherent}Partially coherent candidate limit cycle states}

\begin{figure}[!tb]
    \centering
    \includegraphics[width=0.8\columnwidth]{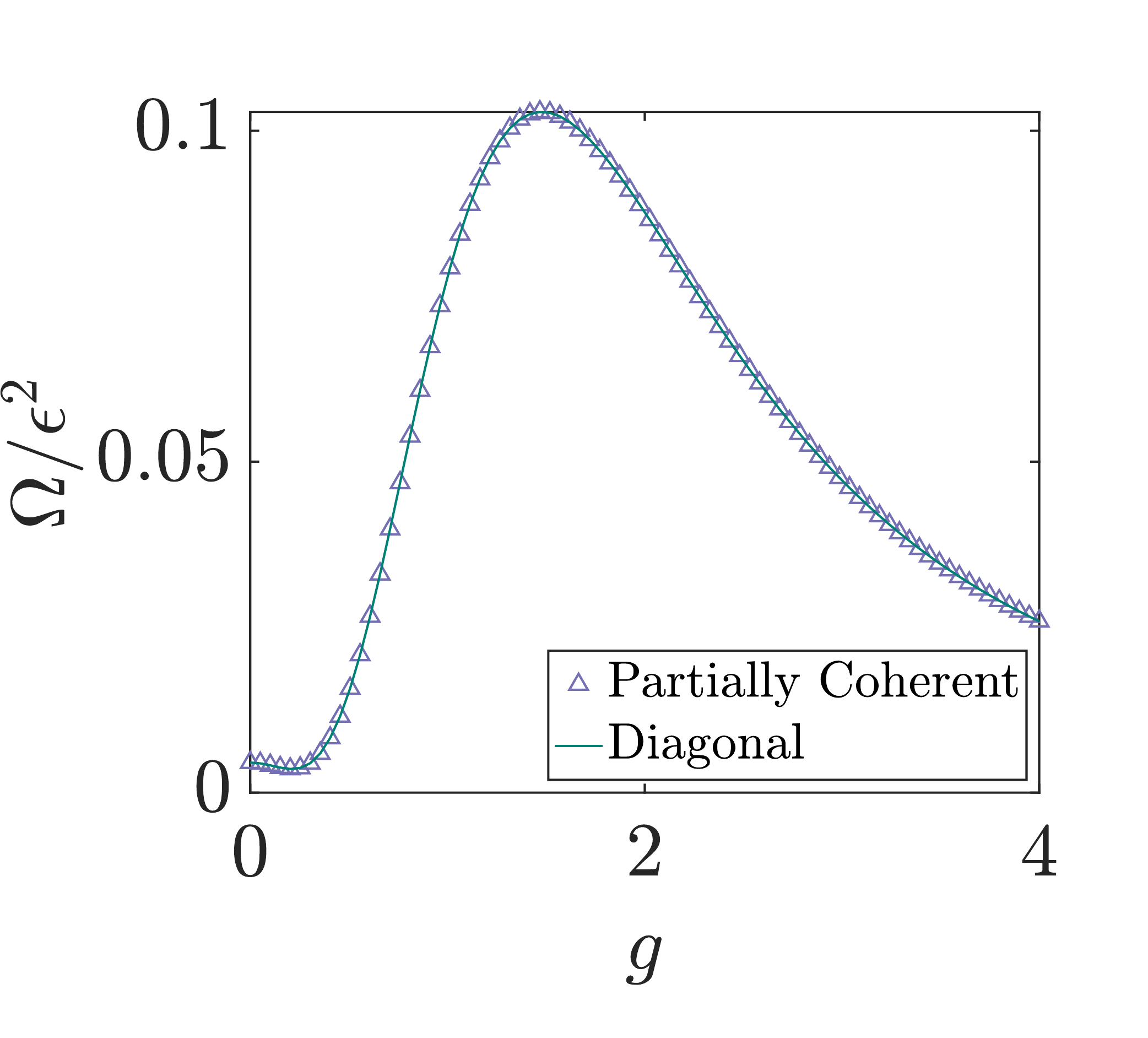}
    \caption{Synchronisation measure ($\Omega/\epsilon^2$) of the composite two-qubit oscillator computed using the diagonal limit cycle state and an optimization over the partially-coherent limit cycle states. Here, we choose $w_A\approx0.55,\gamma_A\approx0.45$ and $w_B\approx0.09,\gamma_B\approx0.91$. Results are obtained numerically.}
    \label{fig:composite-check}
\end{figure}

In Sec.~\ref{sec: metric-composite}, we describe a metric [Eq.~(\ref{eqn:sync-metric})] for studying the synchronization of the two-qubit oscillator, which we use to obtain the results in Sec.~\ref{sec: Synchronisation}. This metric is obtained considering a family of limit cycle states $\Sigma$ that is diagonal in the eigenbasis $\{\ket{\lambda_j}\}$ of the undriven steady state $\hat{\rho}_u$. We note that the $\{\ket{\lambda_j}\}$ are not the eigenstates $\{\ket{E_j}\}$ of the undriven \emph{Hamiltonian}, i.e. Eq.~(\ref{eqn:h_tot}) with $\epsilon=0$. Under such circumstances, Ref.~\cite{jaseem2020generalized} proposes to optimize over a more general family of limit cycle states that allow for partial coherence in the $\{\ket{E_j}\}$ basis, such that the resulting family of states respects the structure of $\hat{\rho}_u$ expressed in this basis. Here, we demonstrate that in our model and for weak driving, the metric~(\ref{eqn:sync-metric}) essentially coincides with the measure obtained by optimization over such partially coherent limit cycle states.

We first note that, in our model, both sets, $\{\ket{\lambda_j}\}$ and $\{\ket{E_j}\}$, are eigenstates of the operator $\hat{S}^z=\hat{S}^z_A+\hat{S}^z_B$. As a result, $\hat{\rho}_u$ can be written in a block-diagonal form in the basis $\{\ket{E_j}\}$, with each block corresponding to a fixed number of total excitations. Under such a situation, $\Sigma$ must be chosen as a set of partially-coherent candidate limit cycle states that account for the intrinsic coherences in each block that are not established by the drive. On the other hand, the matrix elements of the external drive ($\epsilon$ term in Eq.~(\ref{eqn:h_tot})) are block off-diagonal in $\{\ket{E_j}\}$. In other words, $\hat{\rho}_u$ respects a global $U(1)$ symmetry, i.e. it is invariant under unitary transformations of the form $\hat{U}(\varphi) = e^{i\varphi\hat{S}^z}$, while the weak drive, to leading order, only introduces coherences between different blocks associated with this symmetry. As a result, for weak driving, we expect the synchronization measure computed by choosing $\Sigma$ as the set of states diagonal in $\{\lambda_j\}$ (as done in the main text) to coincide with the measure computed using the set of states with partial coherence (as described above) in the $\{\ket{E_j}\}$ basis.

In Fig.~\ref{fig:composite-check}, we demonstrate the excellent agreement between the two approaches. We compute the synchronization measure using partially-coherent candidate limit cycle states via numerical optimization of the limit cycle after imposing the block-diagonal structure in the $\{\ket{E_j}\}$ basis. Indeed, we find that the optimized partially coherent limit cycle state coincides with $\hat{\rho}_\mathrm{diag}$. In the main text, we choose to work with the metric based on diagonal limit cycle states as they are conceptually simpler and more intuitive.

\section{\label{appendix:cumulant-approach} Phase response of individual qubits: Analytical solution}

In this appendix, we outline our procedure to obtain analytical expressions for the phase response measures $\abs{\ev{\hat{S}_j^+}},j=A,B$, using which we rigorously establish the presence of zero-crossing points in Sec.~\ref{sec:Results}. 

The master equation for the system, including dephasing noise on the qudits for a general treatment, is
\begin{eqnarray}
    \label{eq:dephase-master}
    \frac{d\hat{\rho}}{dt} = -i\left[\hat{H}_\mathrm{tot}, \hat{\rho}\right] + \sum_{j={A,B}}\mathfrak{D}[\sqrt{w_j}\hat{S}_j^+]\hat{\rho}\\ \nonumber
    +\sum_{j={A,B}}\mathfrak{D}[\sqrt{\gamma_j}\hat{S}_j^-]\hat{\rho}
    +\sum_{j={A,B}}\mathfrak{D}[\sqrt{2\gamma_\phi}\hat{S}_j^z]\hat{\rho},
\end{eqnarray}
where $\hat{H}_\mathrm{tot}$ is given by Eq.~(\ref{eqn:h_tot}). We treat the drive as a perturbation and expand all observables in orders of $\epsilon$ as
\begin{equation}
    \label{eq: ordereps}
    \braket{\hat{O}} = \braket{\hat{O}}_0 + \epsilon\braket{\hat{O}}_1 +\mathcal{O} (\epsilon ^2).
\end{equation}
At zeroth order in $\epsilon$, the master equation is $U(1)$ symmetric, i.e. it is invariant under the transformation $\hat{S}_{A(B)}^\pm \rightarrow \hat{S}_{A(B)}^\pm e^{\pm i\varphi}$. As a result, only observables that are invariant under this symmetry are non-zero. There are four such quantities, corresponding (at zeroth order) to $\braket{\hat{S}_A^z}_0, \braket{\hat{S}_B^z}_0, \braket{\hat{S}_A^z\hat{S}_B^z}_0, \braket{\hat{S}_A^{+}\hat{S}_B^{-}}_0$. Their equations of motion constitute a set of linear equations given by 

\begin{widetext}
\begin{eqnarray}
\allowdisplaybreaks
    \label{eq:u1-eq}
    &&\mathrm{Re}\left\{i g\braket{\hat{S}_A^{-}\hat{S}_B^{+}}_0\right\}-\Gamma_A\braket{\hat{S}_A^z}_0= -\frac{w_A-\gamma_A}{2} \\ \nonumber
    &&\mathrm{Re}\left\{i g\braket{\hat{S}_A^{+}\hat{S}_B^{-}}_0\right\}-\Gamma_B\braket{\hat{S}_B^z}_0= -\frac{w_B-\gamma_B}{2} \\ \nonumber
    &&\frac{w_B-\gamma_B}{2}\braket{\hat{S}_A^z}_0+\frac{w_A-\gamma_A}{2}\braket{\hat{S}_B^z}_0-(\Gamma_A+\Gamma_B)\braket{\hat{S}_A^z\hat{S}_B^z}_0=0 \\ \nonumber
    &&i\frac{g}{2}\left(\braket{\hat{S}_B^z}_0-\braket{\hat{S}_A^z}_0\right)-\frac{\Gamma_A+\Gamma_B+4\gamma_\phi}{2}\braket{\hat{S}_A^{+}\hat{S}_B^{-}}_0-i\Delta_q\braket{\hat{S}_A^{+}\hat{S}_B^{-}}_0=0,  
\end{eqnarray}
\end{widetext}
where $\Gamma_{A(B)} = w_{A(B)}+\gamma_{A(B)}$. At first order in $\epsilon$,  observables with broken $U(1)$ symmetry acquire a non-zero value. In particular, their equations of motion are `sourced' by the zeroth order $U(1)$ symmetric observables as given below:
\begin{widetext}
\begin{eqnarray}
    \label{eq:nu1-eq}
    &&-ig \braket{\hat{S}_A^z\hat{S}_B^{+}}_1- \frac{\Gamma_A+2\gamma_\phi}{2}\braket{\hat{S}_A^{+}}_1+i\Delta_d\braket{\hat{S}_A^{+}}_1 = i\braket{\hat{S}_A^z}_0 \\ \nonumber
    &&-ig \braket{\hat{S}_A^{+}\hat{S}_B^z}_1- \frac{\Gamma_B+2\gamma_\phi}{2}\braket{\hat{S}_B^{+}}_1+i(\Delta_d+\Delta_q)\braket{\hat{S}_B^{+}}_1 = 0 \\ \nonumber
    &&-i \frac{g}{4}\braket{\hat{S}_B^{+}}_1- \frac{\Gamma_A+2\gamma_\phi}{2}\braket{\hat{S}_A^{+}\hat{S}_B^z}_1+\frac{w_B-\gamma_B}{2}\braket{\hat{S}_A^{+}}_1 
    - \Gamma_B \braket{\hat{S}_A^{+}\hat{S}_B^z}_1+i\Delta_d\braket{\hat{S}_A^{+}\hat{S}_B^z}_1= i\braket{\hat{S}_A^z\hat{S}_B^z}_0 \\ \nonumber
    &&-i\frac{g}{4}\braket{\hat{S}_A^{+}}_1- \frac{\Gamma_B+2\gamma_\phi}{2}\braket{\hat{S}_A^z\hat{S}_B^+}_1
    +\frac{w_A-\gamma_A}{2}\braket{\hat{S}_B^{+}}_1-\Gamma_A \braket{\hat{S}_A^z\hat{S}_B^+}_1+i(\Delta_d+\Delta_q)\braket{\hat{S}_A^z\hat{S}_B^+}_1= -\frac{i}{2}\braket{\hat{S}_A^{-}\hat{S}_B^+}_0 \\ \nonumber
    &&-\frac{\Gamma_A+\Gamma_B+4\gamma_\phi}{2}\braket{\hat{S}_A^{+}\hat{S}_B^+}_1+i (2\Delta_d+\Delta_q)\braket{\hat{S}_A^{+}\hat{S}_B^+}_1 = 0.
\end{eqnarray}
\end{widetext}

Solving these equations, we arrive at analytic expressions for $\ev{\hat{S}_{A}^+},\ev{\hat{S}_{B}^+}$. However, the general form of these expressions are not compact and hence we do not reproduce them here.

\section{Two-qutrit oscillator}
\label{appsec:spin_1}
Our model can be generalised to explore higher dimensional spin systems. In this appendix, we briefly study an oscillator composed of two interacting qutrits that are each coupled to separate thermal baths.

\begin{figure}[!tb]
    \centering
  \includegraphics[width=0.85\columnwidth]{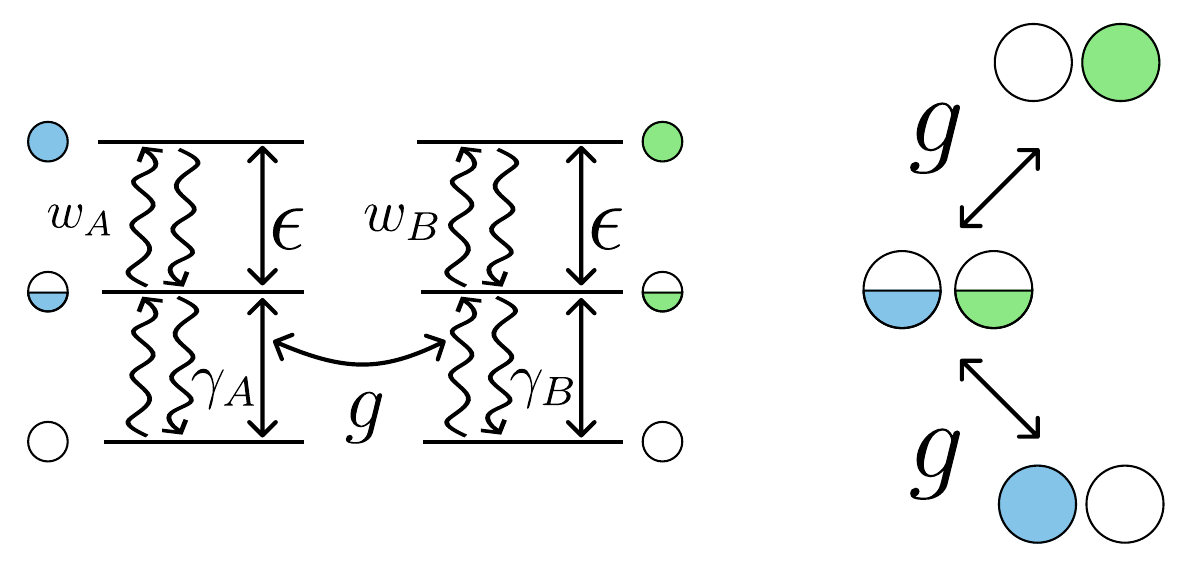}
    \caption{Energy level diagram for the composite two-qutrit oscillator, showing the gain and damping channels and the exchange interaction of the qutrits. The right panel illustrates the exchange interactions. The white, half-colored and colored circles respectively represent the $m=-1,0,1$ states of the two qutrits.}
    \label{fig:modelb}
\end{figure}

\subsection{Phase response of individual qutrits}

\begin{figure}[!tb]
    \centering
    \includegraphics[width=0.8\columnwidth]{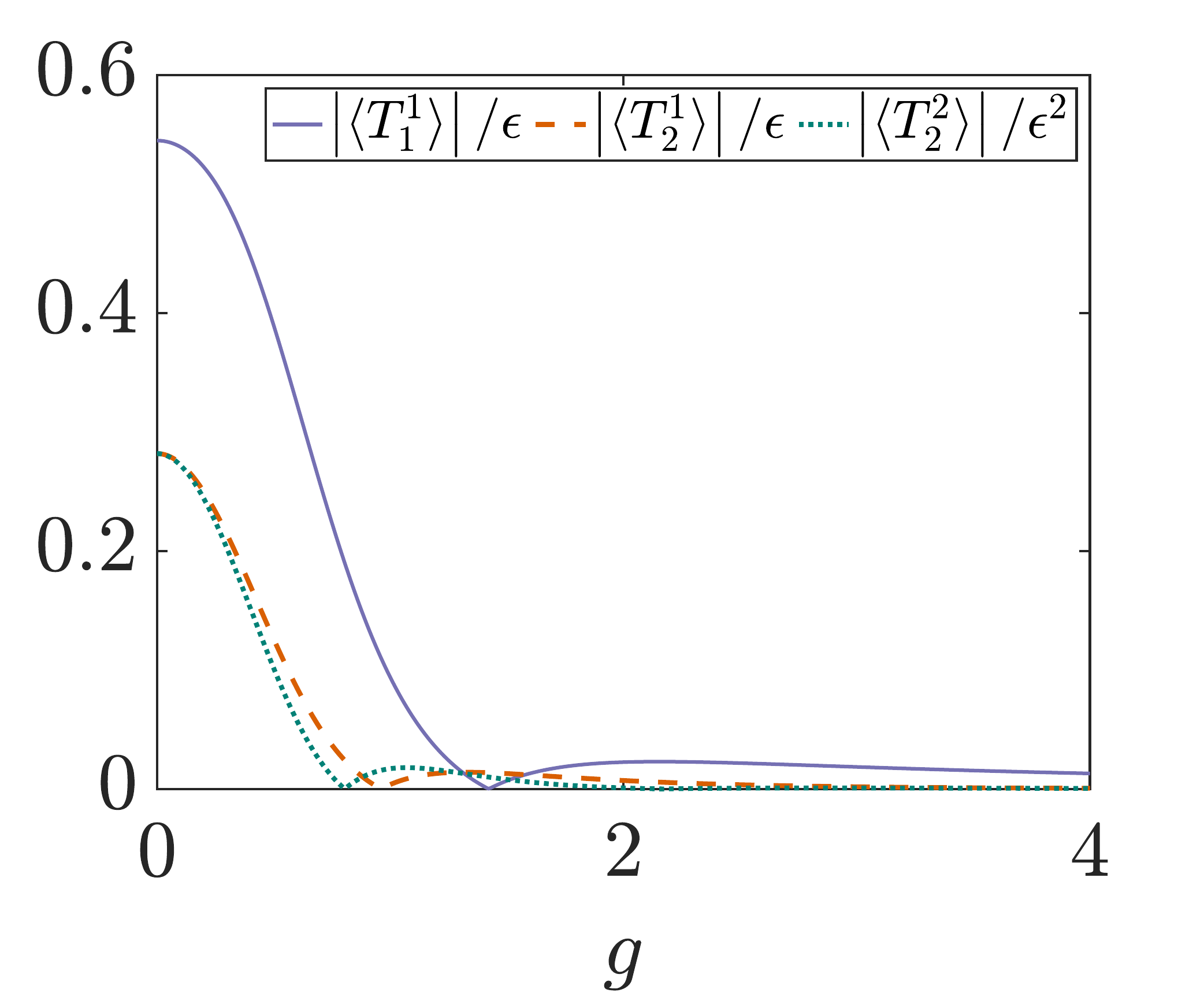}
    \caption{Expectation values of spherical tensor operators for qutrit $A$ versus the qutrit-qutrit coupling strength $g$. Here, we choose $w_A=0.25,\gamma_A=0.75$ and $w_B=0.75,\gamma_B=0.25$, while fixing $w_A+\gamma_A=w_B+\gamma_B=1$. Results are obtained numerically.}
    \label{fig:vanishing-spin1}
\end{figure}

We study the phase response of the constituent qutrits to an external drive applied to one of them using the spherical tensors framework described in Appendix~\ref{appendix:spherical_tensors}. Accordingly, in Fig.~\ref{fig:vanishing-spin1}(a), we plot the quantities $\abs{\braket{\hat{T}_1^1}}, \abs{\braket{\hat{T}_2^1}}$ and $\abs{\braket{\hat{T}_2^2}}$ for qutrit $A$ as a function of the qutrit-qutrit coupling strength and for a fixed set of gain and loss rates for each qutrit. Interestingly, we observe that each of the three quantities undergoes a zero crossing at different coupling strengths. This observation can be interpreted as a generalized blockade effect, where the expectation values of specific spherical tensor multipoles vanish as a result of destructive interference from the coupling to the second qutrit. 

\subsection{Synchronisation of the Composite Two-Qutrit oscillator}

\begin{figure}[!tb]
    \centering
    \includegraphics[width=0.8\columnwidth]{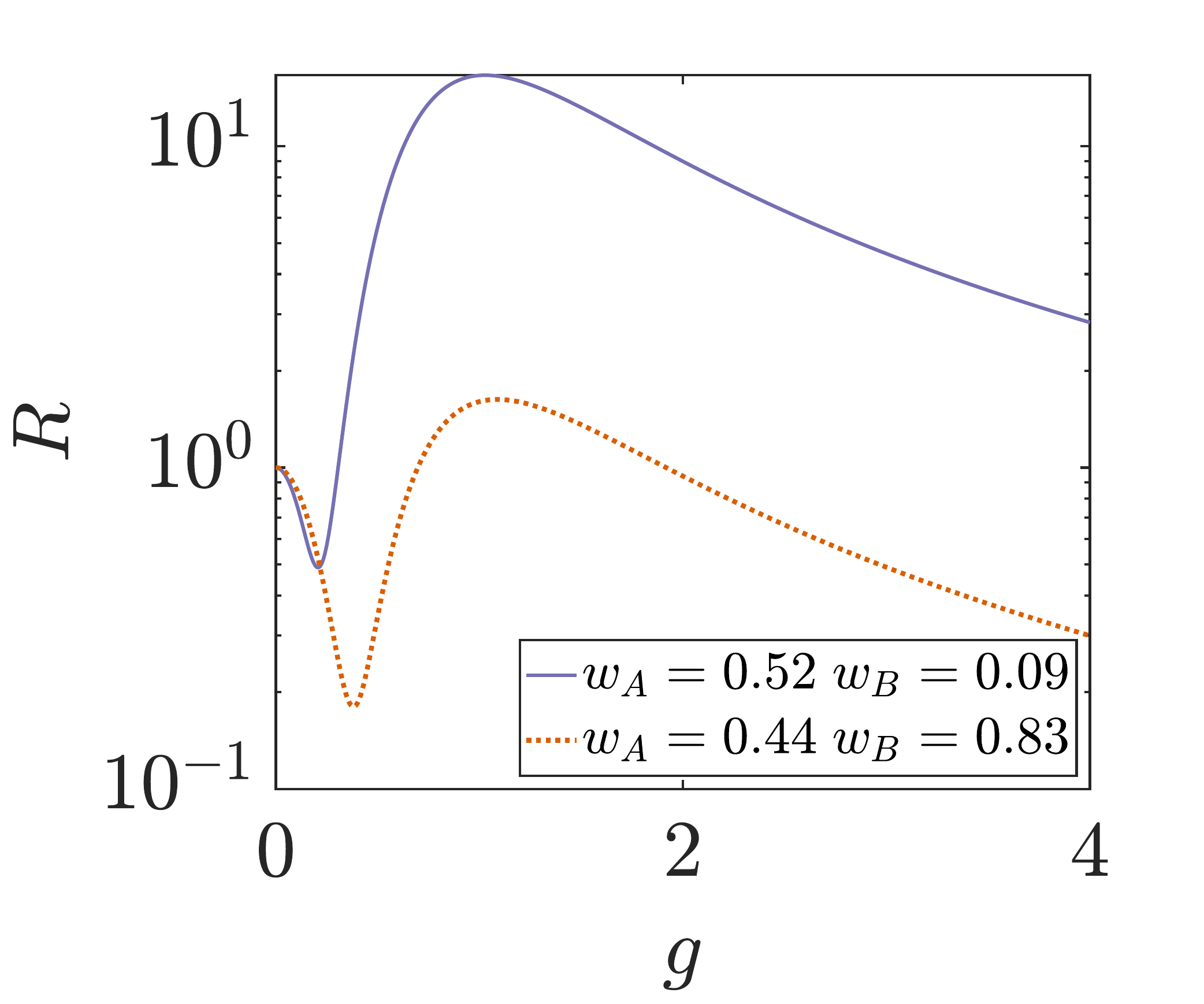}
    \caption{Enhancement and suppression of synchronisation of the composite two-qutrit oscillator. The ratio $R$ [Eq.~(\ref{eqn:r_metric})] versus the coupling strength $g$ for the two-qutrit oscillator and for two different sets of bath parameters, indicated by the $w_A,w_B$ values in the legend. Results are obtained numerically.}
    \label{fig:composite-spin1}
\end{figure}

In  Fig.~\ref{fig:composite-spin1}, we plot the quantity $R$, defined in Eq.~(\ref{eqn:r_metric}), as a function of the qutrit-qutrit coupling strength for two different sets of gain and loss rates for the qutrits. These curves demonstrate that  interactions between the two qutrits can lead to significant enhancement or suppression of synchronization in different parameter regimes, similar to the case of the two-qubit oscillator discussed in the main text. 

\section{Simulations of Proposed cQED Realization}
\label{appendix: details-sc}

In this appendix, we describe the master equation, parameter values and factors considered in choosing these values, for the results presented in Sec.~\ref{sec:cqed}.

The master equation simulation is performed using QuTiP~\cite{johansson2012qutip} for the system depicted in Fig.~\ref{fig:cqed_setup}. We include $3$ levels for each transmon and auxiliary resonator for the simulation and choose to work in a frame that is rotating at the frequencies $\omega_{pA},\omega_{pB}$ of the two-photon `pump' fields, which are denoted by their Rabi frequencies $\Omega_{p_A},\Omega_{p_B}$ in Fig.~\ref{fig:cqed_setup}. The Hamiltonian in such a frame is given by 
\begin{eqnarray}
    \hat{H}_\mathrm{cQED} && = \nonumber \sum_{j={A,B}}(\omega_{qj} - \omega_{pj})\hat{b}^{\dagger}_j\hat{b}_j 
     + \sum_{j={A,B}}\frac{\alpha_j}{2}(\hat{b}^{\dagger}_j\hat{b}_j - 1)\hat{b}^{\dagger}_j\hat{b}_j 
    \\ \nonumber
    && +\sum_{j={A,B}}(\omega_{aj} - \omega_{pj})\hat{a}^{\dagger}_j\hat{a}_j
    \\\nonumber     
    && + \sum_{j={A,B}}g_j(\hat{a}_j\hat{b}^{\dagger}_j + \hat{a}^{\dagger}_j\hat{b}_j) 
    \\ \nonumber 
    && + g(\hat{b}^{\dagger}_A\hat{b}_B e^{i(\omega_{pA}-\omega_{pB})t} + \mathrm{h.c.})
    \\ \nonumber 
    && + \sum_{j={A,B}}\Omega_{pj}(\hat{b}^{\dagger}_j + \hat{b}_j)
    \\ 
    && + \epsilon(\hat{b}_A e^{-i(\omega_{pA} - \omega_{qA})t} + \mathrm{h.c.}).
    \label{eqn:ham_full}
\end{eqnarray}
Here, $\hat{a}_j,\hat{a}_j^\dag$ and $\hat{b}_j,\hat{b}_j^\dag$, $j=A,B$, are the ladder operators for the auxiliary resonators and the transmons respectively. We note that the coupling and drive strengths in this model differ by a factor of two in comparison to the spin model Eq.~(\ref{eqn:h_tot}). The first two lines in Eq.~(\ref{eqn:ham_full}) describe the free Hamiltonian of the transmons and the auxiliary resonators. The third line describes the coupling between the transmons and their respective auxiliary resonators, the fourth line the coupling between the two transmons, the fifth line the two-photon pump on the transmons, and the last line describes the external drive on transmon $A$. The master equation for the full system is given by 
\begin{eqnarray}
    \label{eqn:master_e1_full}
    \frac{d\hat{\rho}}{dt} = -i\left[\hat{H}_\mathrm{cQED}, \hat{\rho}\right] + \sum_{j={A,B}}\mathfrak{D}[\sqrt{\kappa_j}\hat{a}_j]\hat{\rho}\\ \nonumber
    +\sum_{j={A,B}}\mathfrak{D}[\sqrt{\gamma_{0,j}}\hat{b}_j]\hat{\rho}
    +\sum_{j={A,B}}\mathfrak{D}[\sqrt{\gamma_{\phi j}}\hat{b}^{\dagger}_j\hat{b}_j]\hat{\rho}.
\end{eqnarray}
Here, $\kappa_j,\gamma_{0,j}$ describe the decay rates of the resonators and the transmons while $\gamma_{\phi_j}$ describe additional dephasing of the transmons. The values of the parameters entering Eq.~(\ref{eqn:ham_full}) and Eq.~(\ref{eqn:master_e1_full}) are given in Tables~\ref{tab:DeviceParams} and~\ref{tab:DeviceParams2}. These values are experimentally achievable with current technology. 

A number of factors must be carefully considered in choosing parameters for the cQED model, and in order to match its results with the two-qubit oscillator model discussed in Sec.~\ref{sec:Results}. The off-resonant coupling of the $\ket{g}\leftrightarrow\ket{e}$ transition to the auxiliary resonator leads to an additional Purcell decay besides the intrinsic decay channels. The total decay rate $\gamma_j$ and the effective repump rate $w_j$ of each qubit, that are reported in Figs.~\ref{fig:sc-phasecorrelation} and~\ref{fig:sc-enhancement}, are extracted by decoupling the transmons ($g=0$) and fitting the relaxation profiles of the population from an initial state. While the $\ket{g}\leftrightarrow\ket{e}$ transitions of the two transmons must be near-resonance, the corresponding $\ket{e}\leftrightarrow\ket{f}$ transitions must be mismatched in frequency, which will require different anharmonicities for the two transmons. The frequency mismatch ensures that the $\ket{f}\rightarrow\ket{e}$ decay of, say transmon $A$, does not occur through the auxiliary resonator of transmon $B$ or vice versa, by virtue of their coupling. Furthermore, the strength $\epsilon$ of the weak drive cannot be made arbitrarily small since its effects must be discernible in the presence of experimental limitations and residual coherences arising from the pump fields. 

\begin{table}[!tb]
    \centering
    \begin{tabular}{|c|c|c|}
 \hline
Parameter & Symbol & Value \\
 \hline
frequency of qubit A*(B*) & $\omega_{qA}(\omega_{qB})/2\pi$ & 5 GHz \\
 \hline
 frequency of aux A(B) & $\omega_{aA}(\omega_{aB})/2\pi$ & 4.6 GHz \\
 \hline
 anharmonicity of qubit $A$ & $\alpha_A/2\pi$ & 400 MHz \\
 \hline
 anharmonicity of qubit $B$ & $\alpha_B/2\pi$ & 500 MHz \\
 \hline
 qubit-qubit coupling   & $g_{AB}/2\pi$ & 0 - 350 kHz \\
 \hline
 qubit $A$ - aux $A$ coupling & $g_{A}/2\pi$ &  8 MHz \\
 \hline
 qubit $B$ - aux $B$ coupling & $g_{B}/2\pi$ &  4 MHz \\
 \hline
 frequency of qubit $A$ pump  & $\omega_{pA}/2\pi$ & 4.8 GHz \\
 \hline
 frequency of qubit $B$ pump & $\omega_{pB}/2\pi$ & 4.75 GHz \\
 \hline
 decay rate of aux $A$($B$) & $\kappa_{A}(\kappa_{B})/2\pi$ & 60 MHz \\
 \hline
 decay rate of qubit $A$($B$) & $\gamma_{0,A}(\gamma_{0,B})/2\pi$ & 53 kHz \\
 \hline
 dephasing rate of qubit $A$($B$) & $\gamma_{\phi A}(\gamma_{\phi B})/2\pi$ & 53 kHz \\
 \hline
    \end{tabular}    
    \caption{Model parameters. For the quantities marked with a *, the values reported in this Table are approximate and need to be adjusted according to the detunings given in Table~\ref{tab:DeviceParams2}.}
    \label{tab:DeviceParams}
\end{table}

\begin{table}[!tb]
    \centering
    \begin{tabular}{|c|c|c|c|}
\hline
Symbol & Fig \ref{fig:sc-phasecorrelation}(a) &  Fig \ref{fig:sc-phasecorrelation}(b) and Fig \ref{fig:sc-enhancement}(a) & Fig \ref{fig:sc-enhancement}(b) \\
\hline
$\epsilon/2\pi$ & 20 kHz & 20 kHz & 40 kHz   \\
 \hline
 $\Omega_{pA}/2\pi$ & 0.0 MHz & 5.5 MHz & 7 MHz  \\
 \hline
$\Omega_{pB}/2\pi$ & 8.0 MHz & 9.0 MHz & 4.1 MHz \\
 \hline
$\Delta_A/2\pi$ & 160 kHz & 763.3 kHz & 1135 kHz   \\
 \hline
$\Delta_B/2\pi$ & 1013.72 kHz & 1230 kHz & 300 kHz  \\
 \hline
\end{tabular}
\caption{Parameters for different figures. $\Omega_{pA},\Omega_{pB}$ are the Rabi frequencies of the two-photon pump on qubits $A,B$ respectively. $\Delta_A$ and $\Delta_B$ are corrections in the qubit frequency due to the two-photon pump and the auxiliary resonator, such that the shifted frequency of the qubit is given by $\omega_{qj} + \Delta_j,j=A,B$.}
\label{tab:DeviceParams2}
\end{table}

A further, important factor is that the auxiliary resonators and the two-photon pumps introduce shifts to the $\ket{g}\leftrightarrow\ket{e}$ transition frequency of both transmons, which must be compensated by appropriately tuning their frequencies. The dispersive shift of the $\ket{g}\leftrightarrow\ket{e}$ transition frequency arising from the auxiliary resonator is given by  $g_j^2/(\omega_{qj} - \omega_{aj})$.  The shift due to the two-photon pump was calculated by considering the Hamiltonian for the pump acting on the lowest three levels of the transmon. Because of the coupling to the auxiliary resonator, the decay in the third level, given by $\gamma_f = 4g^2/\kappa$\cite{sokolava2021singleatom}, is also included in the Hamiltonian, which, in a frame rotating at the pump frequency takes the form 
\begin{equation}
 H = \begin{bmatrix}
   0 & \Omega_p & 0\\ 
        \Omega_p & \alpha/2 & \sqrt{2}\Omega_p\\
         0 & \sqrt{2}\Omega_p & -i\gamma_f 
   \end{bmatrix}
\end{equation}
where $\Omega_p$ is the two-photon pump strength. The shifted  $\ket{g}\leftrightarrow\ket{e}$ transition frequency is then obtained by diagonalizing this Hamiltonian. The net corrections to the transmon frequencies arising from the auxiliary resonators and the two-photon pumps are listed in Table~\ref{tab:DeviceParams2}.


\begin{thebibliography}{44}%
\makeatletter
\providecommand \@ifxundefined [1]{%
 \@ifx{#1\undefined}
}%
\providecommand \@ifnum [1]{%
 \ifnum #1\expandafter \@firstoftwo
 \else \expandafter \@secondoftwo
 \fi
}%
\providecommand \@ifx [1]{%
 \ifx #1\expandafter \@firstoftwo
 \else \expandafter \@secondoftwo
 \fi
}%
\providecommand \natexlab [1]{#1}%
\providecommand \enquote  [1]{``#1''}%
\providecommand \bibnamefont  [1]{#1}%
\providecommand \bibfnamefont [1]{#1}%
\providecommand \citenamefont [1]{#1}%
\providecommand \href@noop [0]{\@secondoftwo}%
\providecommand \href [0]{\begingroup \@sanitize@url \@href}%
\providecommand \@href[1]{\@@startlink{#1}\@@href}%
\providecommand \@@href[1]{\endgroup#1\@@endlink}%
\providecommand \@sanitize@url [0]{\catcode `\\12\catcode `\$12\catcode `\&12\catcode `\#12\catcode `\^12\catcode `\_12\catcode `\%12\relax}%
\providecommand \@@startlink[1]{}%
\providecommand \@@endlink[0]{}%
\providecommand \url  [0]{\begingroup\@sanitize@url \@url }%
\providecommand \@url [1]{\endgroup\@href {#1}{\urlprefix }}%
\providecommand \urlprefix  [0]{URL }%
\providecommand \Eprint [0]{\href }%
\providecommand \doibase [0]{https://doi.org/}%
\providecommand \selectlanguage [0]{\@gobble}%
\providecommand \bibinfo  [0]{\@secondoftwo}%
\providecommand \bibfield  [0]{\@secondoftwo}%
\providecommand \translation [1]{[#1]}%
\providecommand \BibitemOpen [0]{}%
\providecommand \bibitemStop [0]{}%
\providecommand \bibitemNoStop [0]{.\EOS\space}%
\providecommand \EOS [0]{\spacefactor3000\relax}%
\providecommand \BibitemShut  [1]{\csname bibitem#1\endcsname}%
\let\auto@bib@innerbib\@empty
\bibitem [{\citenamefont {Pikovsky}\ \emph {et~al.}(2001)\citenamefont {Pikovsky}, \citenamefont {Rosenblum},\ and\ \citenamefont {Kurths}}]{pikovsky2001universal}%
  \BibitemOpen
  \bibfield  {author} {\bibinfo {author} {\bibfnamefont {A.}~\bibnamefont {Pikovsky}}, \bibinfo {author} {\bibfnamefont {M.}~\bibnamefont {Rosenblum}},\ and\ \bibinfo {author} {\bibfnamefont {J.}~\bibnamefont {Kurths}},\ }\bibfield  {title} {\bibinfo {title} {Synchronization: A universal concept in nonlinear sciences},\ }\href@noop {} {\bibfield  {journal} {\bibinfo  {journal} {Self}\ }\textbf {\bibinfo {volume} {2}},\ \bibinfo {pages} {3} (\bibinfo {year} {2001})}\BibitemShut {NoStop}%
\bibitem [{\citenamefont {Lee}\ and\ \citenamefont {Sadeghpour}(2013)}]{lee2013quantum}%
  \BibitemOpen
  \bibfield  {author} {\bibinfo {author} {\bibfnamefont {T.~E.}\ \bibnamefont {Lee}}\ and\ \bibinfo {author} {\bibfnamefont {H.}~\bibnamefont {Sadeghpour}},\ }\bibfield  {title} {\bibinfo {title} {Quantum synchronization of quantum van der pol oscillators with trapped ions},\ }\href@noop {} {\bibfield  {journal} {\bibinfo  {journal} {Physical review letters}\ }\textbf {\bibinfo {volume} {111}},\ \bibinfo {pages} {234101} (\bibinfo {year} {2013})}\BibitemShut {NoStop}%
\bibitem [{\citenamefont {Walter}\ \emph {et~al.}(2014)\citenamefont {Walter}, \citenamefont {Nunnenkamp},\ and\ \citenamefont {Bruder}}]{walter2014quantum}%
  \BibitemOpen
  \bibfield  {author} {\bibinfo {author} {\bibfnamefont {S.}~\bibnamefont {Walter}}, \bibinfo {author} {\bibfnamefont {A.}~\bibnamefont {Nunnenkamp}},\ and\ \bibinfo {author} {\bibfnamefont {C.}~\bibnamefont {Bruder}},\ }\bibfield  {title} {\bibinfo {title} {Quantum synchronization of a driven self-sustained oscillator},\ }\href@noop {} {\bibfield  {journal} {\bibinfo  {journal} {Physical review letters}\ }\textbf {\bibinfo {volume} {112}},\ \bibinfo {pages} {094102} (\bibinfo {year} {2014})}\BibitemShut {NoStop}%
\bibitem [{\citenamefont {Lee}\ \emph {et~al.}(2014)\citenamefont {Lee}, \citenamefont {Chan},\ and\ \citenamefont {Wang}}]{lee2014entanglement}%
  \BibitemOpen
  \bibfield  {author} {\bibinfo {author} {\bibfnamefont {T.~E.}\ \bibnamefont {Lee}}, \bibinfo {author} {\bibfnamefont {C.-K.}\ \bibnamefont {Chan}},\ and\ \bibinfo {author} {\bibfnamefont {S.}~\bibnamefont {Wang}},\ }\bibfield  {title} {\bibinfo {title} {Entanglement tongue and quantum synchronization of disordered oscillators},\ }\href@noop {} {\bibfield  {journal} {\bibinfo  {journal} {Physical Review E}\ }\textbf {\bibinfo {volume} {89}},\ \bibinfo {pages} {022913} (\bibinfo {year} {2014})}\BibitemShut {NoStop}%
\bibitem [{\citenamefont {Sonar}\ \emph {et~al.}(2018)\citenamefont {Sonar}, \citenamefont {Hajdu{\v{s}}ek}, \citenamefont {Mukherjee}, \citenamefont {Fazio}, \citenamefont {Vedral}, \citenamefont {Vinjanampathy},\ and\ \citenamefont {Kwek}}]{sonar2018squeezing}%
  \BibitemOpen
  \bibfield  {author} {\bibinfo {author} {\bibfnamefont {S.}~\bibnamefont {Sonar}}, \bibinfo {author} {\bibfnamefont {M.}~\bibnamefont {Hajdu{\v{s}}ek}}, \bibinfo {author} {\bibfnamefont {M.}~\bibnamefont {Mukherjee}}, \bibinfo {author} {\bibfnamefont {R.}~\bibnamefont {Fazio}}, \bibinfo {author} {\bibfnamefont {V.}~\bibnamefont {Vedral}}, \bibinfo {author} {\bibfnamefont {S.}~\bibnamefont {Vinjanampathy}},\ and\ \bibinfo {author} {\bibfnamefont {L.-C.}\ \bibnamefont {Kwek}},\ }\bibfield  {title} {\bibinfo {title} {Squeezing enhances quantum synchronization},\ }\href@noop {} {\bibfield  {journal} {\bibinfo  {journal} {Physical review letters}\ }\textbf {\bibinfo {volume} {120}},\ \bibinfo {pages} {163601} (\bibinfo {year} {2018})}\BibitemShut {NoStop}%
\bibitem [{\citenamefont {Roulet}\ and\ \citenamefont {Bruder}(2018{\natexlab{a}})}]{roulet2018quantum}%
  \BibitemOpen
  \bibfield  {author} {\bibinfo {author} {\bibfnamefont {A.}~\bibnamefont {Roulet}}\ and\ \bibinfo {author} {\bibfnamefont {C.}~\bibnamefont {Bruder}},\ }\bibfield  {title} {\bibinfo {title} {Quantum synchronization and entanglement generation},\ }\href@noop {} {\bibfield  {journal} {\bibinfo  {journal} {Physical review letters}\ }\textbf {\bibinfo {volume} {121}},\ \bibinfo {pages} {063601} (\bibinfo {year} {2018}{\natexlab{a}})}\BibitemShut {NoStop}%
\bibitem [{\citenamefont {Roulet}\ and\ \citenamefont {Bruder}(2018{\natexlab{b}})}]{roulet2018synchronizing}%
  \BibitemOpen
  \bibfield  {author} {\bibinfo {author} {\bibfnamefont {A.}~\bibnamefont {Roulet}}\ and\ \bibinfo {author} {\bibfnamefont {C.}~\bibnamefont {Bruder}},\ }\bibfield  {title} {\bibinfo {title} {Synchronizing the smallest possible system},\ }\href@noop {} {\bibfield  {journal} {\bibinfo  {journal} {Physical review letters}\ }\textbf {\bibinfo {volume} {121}},\ \bibinfo {pages} {053601} (\bibinfo {year} {2018}{\natexlab{b}})}\BibitemShut {NoStop}%
\bibitem [{\citenamefont {Koppenh\"ofer}\ and\ \citenamefont {Roulet}(2019)}]{koppenhofer2019PRA}%
  \BibitemOpen
  \bibfield  {author} {\bibinfo {author} {\bibfnamefont {M.}~\bibnamefont {Koppenh\"ofer}}\ and\ \bibinfo {author} {\bibfnamefont {A.}~\bibnamefont {Roulet}},\ }\bibfield  {title} {\bibinfo {title} {Optimal synchronization deep in the quantum regime: Resource and fundamental limit},\ }\href {https://doi.org/10.1103/PhysRevA.99.043804} {\bibfield  {journal} {\bibinfo  {journal} {Phys. Rev. A}\ }\textbf {\bibinfo {volume} {99}},\ \bibinfo {pages} {043804} (\bibinfo {year} {2019})}\BibitemShut {NoStop}%
\bibitem [{\citenamefont {Tan}\ \emph {et~al.}(2022)\citenamefont {Tan}, \citenamefont {Bruder},\ and\ \citenamefont {Koppenh{\"{o}}fer}}]{tan2022quantum}%
  \BibitemOpen
  \bibfield  {author} {\bibinfo {author} {\bibfnamefont {R.}~\bibnamefont {Tan}}, \bibinfo {author} {\bibfnamefont {C.}~\bibnamefont {Bruder}},\ and\ \bibinfo {author} {\bibfnamefont {M.}~\bibnamefont {Koppenh{\"{o}}fer}},\ }\bibfield  {title} {\bibinfo {title} {Half-integer vs. integer effects in quantum synchronization of spin systems},\ }\href {https://doi.org/10.22331/q-2022-12-29-885} {\bibfield  {journal} {\bibinfo  {journal} {{Quantum}}\ }\textbf {\bibinfo {volume} {6}},\ \bibinfo {pages} {885} (\bibinfo {year} {2022})}\BibitemShut {NoStop}%
\bibitem [{\citenamefont {L{\"o}rch}\ \emph {et~al.}(2017)\citenamefont {L{\"o}rch}, \citenamefont {Nigg}, \citenamefont {Nunnenkamp}, \citenamefont {Tiwari},\ and\ \citenamefont {Bruder}}]{lorch2017quantum}%
  \BibitemOpen
  \bibfield  {author} {\bibinfo {author} {\bibfnamefont {N.}~\bibnamefont {L{\"o}rch}}, \bibinfo {author} {\bibfnamefont {S.~E.}\ \bibnamefont {Nigg}}, \bibinfo {author} {\bibfnamefont {A.}~\bibnamefont {Nunnenkamp}}, \bibinfo {author} {\bibfnamefont {R.~P.}\ \bibnamefont {Tiwari}},\ and\ \bibinfo {author} {\bibfnamefont {C.}~\bibnamefont {Bruder}},\ }\bibfield  {title} {\bibinfo {title} {Quantum synchronization blockade: Energy quantization hinders synchronization of identical oscillators},\ }\href@noop {} {\bibfield  {journal} {\bibinfo  {journal} {Physical Review Letters}\ }\textbf {\bibinfo {volume} {118}},\ \bibinfo {pages} {243602} (\bibinfo {year} {2017})}\BibitemShut {NoStop}%
\bibitem [{\citenamefont {Nigg}(2018)}]{nigg2018observing}%
  \BibitemOpen
  \bibfield  {author} {\bibinfo {author} {\bibfnamefont {S.~E.}\ \bibnamefont {Nigg}},\ }\bibfield  {title} {\bibinfo {title} {Observing quantum synchronization blockade in circuit quantum electrodynamics},\ }\href@noop {} {\bibfield  {journal} {\bibinfo  {journal} {Physical Review A}\ }\textbf {\bibinfo {volume} {97}},\ \bibinfo {pages} {013811} (\bibinfo {year} {2018})}\BibitemShut {NoStop}%
\bibitem [{\citenamefont {Solanki}\ \emph {et~al.}(2022)\citenamefont {Solanki}, \citenamefont {Mehdi}, \citenamefont {Hajdu{\v{s}}ek},\ and\ \citenamefont {Vinjanampathy}}]{solanki2022symmetries}%
  \BibitemOpen
  \bibfield  {author} {\bibinfo {author} {\bibfnamefont {P.}~\bibnamefont {Solanki}}, \bibinfo {author} {\bibfnamefont {F.~M.}\ \bibnamefont {Mehdi}}, \bibinfo {author} {\bibfnamefont {M.}~\bibnamefont {Hajdu{\v{s}}ek}},\ and\ \bibinfo {author} {\bibfnamefont {S.}~\bibnamefont {Vinjanampathy}},\ }\bibfield  {title} {\bibinfo {title} {Symmetries and synchronization blockade},\ }\href@noop {} {\bibfield  {journal} {\bibinfo  {journal} {arXiv preprint arXiv:2212.09388}\ } (\bibinfo {year} {2022})}\BibitemShut {NoStop}%
\bibitem [{\citenamefont {Laskar}\ \emph {et~al.}(2020)\citenamefont {Laskar}, \citenamefont {Adhikary}, \citenamefont {Mondal}, \citenamefont {Katiyar}, \citenamefont {Vinjanampathy},\ and\ \citenamefont {Ghosh}}]{laskar2020observation}%
  \BibitemOpen
  \bibfield  {author} {\bibinfo {author} {\bibfnamefont {A.~W.}\ \bibnamefont {Laskar}}, \bibinfo {author} {\bibfnamefont {P.}~\bibnamefont {Adhikary}}, \bibinfo {author} {\bibfnamefont {S.}~\bibnamefont {Mondal}}, \bibinfo {author} {\bibfnamefont {P.}~\bibnamefont {Katiyar}}, \bibinfo {author} {\bibfnamefont {S.}~\bibnamefont {Vinjanampathy}},\ and\ \bibinfo {author} {\bibfnamefont {S.}~\bibnamefont {Ghosh}},\ }\bibfield  {title} {\bibinfo {title} {Observation of quantum phase synchronization in spin-1 atoms},\ }\href@noop {} {\bibfield  {journal} {\bibinfo  {journal} {Physical review letters}\ }\textbf {\bibinfo {volume} {125}},\ \bibinfo {pages} {013601} (\bibinfo {year} {2020})}\BibitemShut {NoStop}%
\bibitem [{\citenamefont {Krithika}\ \emph {et~al.}(2022)\citenamefont {Krithika}, \citenamefont {Solanki}, \citenamefont {Vinjanampathy},\ and\ \citenamefont {Mahesh}}]{krithika2022observation}%
  \BibitemOpen
  \bibfield  {author} {\bibinfo {author} {\bibfnamefont {V.}~\bibnamefont {Krithika}}, \bibinfo {author} {\bibfnamefont {P.}~\bibnamefont {Solanki}}, \bibinfo {author} {\bibfnamefont {S.}~\bibnamefont {Vinjanampathy}},\ and\ \bibinfo {author} {\bibfnamefont {T.}~\bibnamefont {Mahesh}},\ }\bibfield  {title} {\bibinfo {title} {Observation of quantum phase synchronization in a nuclear-spin system},\ }\href@noop {} {\bibfield  {journal} {\bibinfo  {journal} {Physical Review A}\ }\textbf {\bibinfo {volume} {105}},\ \bibinfo {pages} {062206} (\bibinfo {year} {2022})}\BibitemShut {NoStop}%
\bibitem [{\citenamefont {Koppenh{\"o}fer}\ \emph {et~al.}(2020)\citenamefont {Koppenh{\"o}fer}, \citenamefont {Bruder},\ and\ \citenamefont {Roulet}}]{koppenhofer2020quantum}%
  \BibitemOpen
  \bibfield  {author} {\bibinfo {author} {\bibfnamefont {M.}~\bibnamefont {Koppenh{\"o}fer}}, \bibinfo {author} {\bibfnamefont {C.}~\bibnamefont {Bruder}},\ and\ \bibinfo {author} {\bibfnamefont {A.}~\bibnamefont {Roulet}},\ }\bibfield  {title} {\bibinfo {title} {Quantum synchronization on the ibm q system},\ }\href@noop {} {\bibfield  {journal} {\bibinfo  {journal} {Physical Review Research}\ }\textbf {\bibinfo {volume} {2}},\ \bibinfo {pages} {023026} (\bibinfo {year} {2020})}\BibitemShut {NoStop}%
\bibitem [{\citenamefont {Jaseem}\ \emph {et~al.}(2020{\natexlab{a}})\citenamefont {Jaseem}, \citenamefont {Hajdu{\v{s}}ek}, \citenamefont {Solanki}, \citenamefont {Kwek}, \citenamefont {Fazio},\ and\ \citenamefont {Vinjanampathy}}]{jaseem2020generalized}%
  \BibitemOpen
  \bibfield  {author} {\bibinfo {author} {\bibfnamefont {N.}~\bibnamefont {Jaseem}}, \bibinfo {author} {\bibfnamefont {M.}~\bibnamefont {Hajdu{\v{s}}ek}}, \bibinfo {author} {\bibfnamefont {P.}~\bibnamefont {Solanki}}, \bibinfo {author} {\bibfnamefont {L.-C.}\ \bibnamefont {Kwek}}, \bibinfo {author} {\bibfnamefont {R.}~\bibnamefont {Fazio}},\ and\ \bibinfo {author} {\bibfnamefont {S.}~\bibnamefont {Vinjanampathy}},\ }\bibfield  {title} {\bibinfo {title} {Generalized measure of quantum synchronization},\ }\href@noop {} {\bibfield  {journal} {\bibinfo  {journal} {Physical Review Research}\ }\textbf {\bibinfo {volume} {2}},\ \bibinfo {pages} {043287} (\bibinfo {year} {2020}{\natexlab{a}})}\BibitemShut {NoStop}%
\bibitem [{\citenamefont {Ameri}\ \emph {et~al.}(2015)\citenamefont {Ameri}, \citenamefont {Eghbali-Arani}, \citenamefont {Mari}, \citenamefont {Farace}, \citenamefont {Kheirandish}, \citenamefont {Giovannetti},\ and\ \citenamefont {Fazio}}]{ameri2015PRA}%
  \BibitemOpen
  \bibfield  {author} {\bibinfo {author} {\bibfnamefont {V.}~\bibnamefont {Ameri}}, \bibinfo {author} {\bibfnamefont {M.}~\bibnamefont {Eghbali-Arani}}, \bibinfo {author} {\bibfnamefont {A.}~\bibnamefont {Mari}}, \bibinfo {author} {\bibfnamefont {A.}~\bibnamefont {Farace}}, \bibinfo {author} {\bibfnamefont {F.}~\bibnamefont {Kheirandish}}, \bibinfo {author} {\bibfnamefont {V.}~\bibnamefont {Giovannetti}},\ and\ \bibinfo {author} {\bibfnamefont {R.}~\bibnamefont {Fazio}},\ }\bibfield  {title} {\bibinfo {title} {Mutual information as an order parameter for quantum synchronization},\ }\href {https://doi.org/10.1103/PhysRevA.91.012301} {\bibfield  {journal} {\bibinfo  {journal} {Phys. Rev. A}\ }\textbf {\bibinfo {volume} {91}},\ \bibinfo {pages} {012301} (\bibinfo {year} {2015})}\BibitemShut {NoStop}%
\bibitem [{\citenamefont {Shen}\ \emph {et~al.}(2023)\citenamefont {Shen}, \citenamefont {Mok}, \citenamefont {Noh}, \citenamefont {Liu}, \citenamefont {Kwek}, \citenamefont {Fan},\ and\ \citenamefont {Chia}}]{shen2023quantum}%
  \BibitemOpen
  \bibfield  {author} {\bibinfo {author} {\bibfnamefont {Y.}~\bibnamefont {Shen}}, \bibinfo {author} {\bibfnamefont {W.-K.}\ \bibnamefont {Mok}}, \bibinfo {author} {\bibfnamefont {C.}~\bibnamefont {Noh}}, \bibinfo {author} {\bibfnamefont {A.~Q.}\ \bibnamefont {Liu}}, \bibinfo {author} {\bibfnamefont {L.-C.}\ \bibnamefont {Kwek}}, \bibinfo {author} {\bibfnamefont {W.}~\bibnamefont {Fan}},\ and\ \bibinfo {author} {\bibfnamefont {A.}~\bibnamefont {Chia}},\ }\bibfield  {title} {\bibinfo {title} {Quantum synchronization effects induced by strong nonlinearities},\ }\href@noop {} {\bibfield  {journal} {\bibinfo  {journal} {arXiv preprint arXiv:2301.02948}\ } (\bibinfo {year} {2023})}\BibitemShut {NoStop}%
\bibitem [{\citenamefont {Buča}\ \emph {et~al.}(2022)\citenamefont {Buča}, \citenamefont {Booker},\ and\ \citenamefont {Jaksch}}]{buca2022SciPost}%
  \BibitemOpen
  \bibfield  {author} {\bibinfo {author} {\bibfnamefont {B.}~\bibnamefont {Buča}}, \bibinfo {author} {\bibfnamefont {C.}~\bibnamefont {Booker}},\ and\ \bibinfo {author} {\bibfnamefont {D.}~\bibnamefont {Jaksch}},\ }\bibfield  {title} {\bibinfo {title} {{Algebraic theory of quantum synchronization and limit cycles under dissipation}},\ }\href {https://doi.org/10.21468/SciPostPhys.12.3.097} {\bibfield  {journal} {\bibinfo  {journal} {SciPost Phys.}\ }\textbf {\bibinfo {volume} {12}},\ \bibinfo {pages} {097} (\bibinfo {year} {2022})}\BibitemShut {NoStop}%
\bibitem [{\citenamefont {Tindall}\ \emph {et~al.}(2020)\citenamefont {Tindall}, \citenamefont {Muñoz}, \citenamefont {Buča},\ and\ \citenamefont {Jaksch}}]{Tindall_2020}%
  \BibitemOpen
  \bibfield  {author} {\bibinfo {author} {\bibfnamefont {J.}~\bibnamefont {Tindall}}, \bibinfo {author} {\bibfnamefont {C.~S.}\ \bibnamefont {Muñoz}}, \bibinfo {author} {\bibfnamefont {B.}~\bibnamefont {Buča}},\ and\ \bibinfo {author} {\bibfnamefont {D.}~\bibnamefont {Jaksch}},\ }\bibfield  {title} {\bibinfo {title} {Quantum synchronisation enabled by dynamical symmetries and dissipation},\ }\href {https://doi.org/10.1088/1367-2630/ab60f5} {\bibfield  {journal} {\bibinfo  {journal} {New Journal of Physics}\ }\textbf {\bibinfo {volume} {22}},\ \bibinfo {pages} {013026} (\bibinfo {year} {2020})}\BibitemShut {NoStop}%
\bibitem [{\citenamefont {Murtadho}\ \emph {et~al.}(2023{\natexlab{a}})\citenamefont {Murtadho}, \citenamefont {Vinjanampathy},\ and\ \citenamefont {Thingna}}]{murtadho2023PRL}%
  \BibitemOpen
  \bibfield  {author} {\bibinfo {author} {\bibfnamefont {T.}~\bibnamefont {Murtadho}}, \bibinfo {author} {\bibfnamefont {S.}~\bibnamefont {Vinjanampathy}},\ and\ \bibinfo {author} {\bibfnamefont {J.}~\bibnamefont {Thingna}},\ }\bibfield  {title} {\bibinfo {title} {Cooperation and competition in synchronous open quantum systems},\ }\href {https://doi.org/10.1103/PhysRevLett.131.030401} {\bibfield  {journal} {\bibinfo  {journal} {Phys. Rev. Lett.}\ }\textbf {\bibinfo {volume} {131}},\ \bibinfo {pages} {030401} (\bibinfo {year} {2023}{\natexlab{a}})}\BibitemShut {NoStop}%
\bibitem [{\citenamefont {Scala}\ \emph {et~al.}(2011)\citenamefont {Scala}, \citenamefont {Migliore}, \citenamefont {Messina},\ and\ \citenamefont {S{\'a}nchez-Soto}}]{scala2011EPJD}%
  \BibitemOpen
  \bibfield  {author} {\bibinfo {author} {\bibfnamefont {M.}~\bibnamefont {Scala}}, \bibinfo {author} {\bibfnamefont {R.}~\bibnamefont {Migliore}}, \bibinfo {author} {\bibfnamefont {A.}~\bibnamefont {Messina}},\ and\ \bibinfo {author} {\bibfnamefont {L.~L.}\ \bibnamefont {S{\'a}nchez-Soto}},\ }\bibfield  {title} {\bibinfo {title} {Robust stationary entanglement of two coupled qubits inindependent environments},\ }\href {https://doi.org/10.1140/epjd/e2010-00230-5} {\bibfield  {journal} {\bibinfo  {journal} {The European Physical Journal D}\ }\textbf {\bibinfo {volume} {61}},\ \bibinfo {pages} {199} (\bibinfo {year} {2011})}\BibitemShut {NoStop}%
\bibitem [{\citenamefont {Brask}\ \emph {et~al.}(2015)\citenamefont {Brask}, \citenamefont {Haack}, \citenamefont {Brunner},\ and\ \citenamefont {Huber}}]{brask2015NJP}%
  \BibitemOpen
  \bibfield  {author} {\bibinfo {author} {\bibfnamefont {J.~B.}\ \bibnamefont {Brask}}, \bibinfo {author} {\bibfnamefont {G.}~\bibnamefont {Haack}}, \bibinfo {author} {\bibfnamefont {N.}~\bibnamefont {Brunner}},\ and\ \bibinfo {author} {\bibfnamefont {M.}~\bibnamefont {Huber}},\ }\bibfield  {title} {\bibinfo {title} {Autonomous quantum thermal machine for generating steady-state entanglement},\ }\href {https://doi.org/10.1088/1367-2630/17/11/113029} {\bibfield  {journal} {\bibinfo  {journal} {New Journal of Physics}\ }\textbf {\bibinfo {volume} {17}},\ \bibinfo {pages} {113029} (\bibinfo {year} {2015})}\BibitemShut {NoStop}%
\bibitem [{\citenamefont {Hofer}\ \emph {et~al.}(2017)\citenamefont {Hofer}, \citenamefont {Perarnau-Llobet}, \citenamefont {Miranda}, \citenamefont {Haack}, \citenamefont {Silva}, \citenamefont {Brask},\ and\ \citenamefont {Brunner}}]{hofer2017NJP}%
  \BibitemOpen
  \bibfield  {author} {\bibinfo {author} {\bibfnamefont {P.~P.}\ \bibnamefont {Hofer}}, \bibinfo {author} {\bibfnamefont {M.}~\bibnamefont {Perarnau-Llobet}}, \bibinfo {author} {\bibfnamefont {L.~D.~M.}\ \bibnamefont {Miranda}}, \bibinfo {author} {\bibfnamefont {G.}~\bibnamefont {Haack}}, \bibinfo {author} {\bibfnamefont {R.}~\bibnamefont {Silva}}, \bibinfo {author} {\bibfnamefont {J.~B.}\ \bibnamefont {Brask}},\ and\ \bibinfo {author} {\bibfnamefont {N.}~\bibnamefont {Brunner}},\ }\bibfield  {title} {\bibinfo {title} {Markovian master equations for quantum thermal machines: local versus global approach},\ }\href {https://doi.org/10.1088/1367-2630/aa964f} {\bibfield  {journal} {\bibinfo  {journal} {New Journal of Physics}\ }\textbf {\bibinfo {volume} {19}},\ \bibinfo {pages} {123037} (\bibinfo {year} {2017})}\BibitemShut {NoStop}%
\bibitem [{\citenamefont {Cattaneo}\ \emph {et~al.}(2019)\citenamefont {Cattaneo}, \citenamefont {Giorgi}, \citenamefont {Maniscalco},\ and\ \citenamefont {Zambrini}}]{cattaneo2019NJP}%
  \BibitemOpen
  \bibfield  {author} {\bibinfo {author} {\bibfnamefont {M.}~\bibnamefont {Cattaneo}}, \bibinfo {author} {\bibfnamefont {G.~L.}\ \bibnamefont {Giorgi}}, \bibinfo {author} {\bibfnamefont {S.}~\bibnamefont {Maniscalco}},\ and\ \bibinfo {author} {\bibfnamefont {R.}~\bibnamefont {Zambrini}},\ }\bibfield  {title} {\bibinfo {title} {Local versus global master equation with common and separate baths: superiority of the global approach in partial secular approximation},\ }\href {https://doi.org/10.1088/1367-2630/ab54ac} {\bibfield  {journal} {\bibinfo  {journal} {New Journal of Physics}\ }\textbf {\bibinfo {volume} {21}},\ \bibinfo {pages} {113045} (\bibinfo {year} {2019})}\BibitemShut {NoStop}%
\bibitem [{\citenamefont {Sokolova}\ \emph {et~al.}(2021)\citenamefont {Sokolova}, \citenamefont {Fedorov}, \citenamefont {Il'ichev},\ and\ \citenamefont {Astafiev}}]{sokolava2021singleatom}%
  \BibitemOpen
  \bibfield  {author} {\bibinfo {author} {\bibfnamefont {A.~A.}\ \bibnamefont {Sokolova}}, \bibinfo {author} {\bibfnamefont {G.~P.}\ \bibnamefont {Fedorov}}, \bibinfo {author} {\bibfnamefont {E.~V.}\ \bibnamefont {Il'ichev}},\ and\ \bibinfo {author} {\bibfnamefont {O.~V.}\ \bibnamefont {Astafiev}},\ }\bibfield  {title} {\bibinfo {title} {Single-atom maser with an engineered circuit for population inversion},\ }\href {https://doi.org/10.1103/PhysRevA.103.013718} {\bibfield  {journal} {\bibinfo  {journal} {Phys. Rev. A}\ }\textbf {\bibinfo {volume} {103}},\ \bibinfo {pages} {013718} (\bibinfo {year} {2021})}\BibitemShut {NoStop}%
\bibitem [{\citenamefont {Sokolova}\ \emph {et~al.}(2023)\citenamefont {Sokolova}, \citenamefont {Kalacheva}, \citenamefont {Fedorov},\ and\ \citenamefont {Astafiev}}]{sokolova2023PRA}%
  \BibitemOpen
  \bibfield  {author} {\bibinfo {author} {\bibfnamefont {A.~A.}\ \bibnamefont {Sokolova}}, \bibinfo {author} {\bibfnamefont {D.~A.}\ \bibnamefont {Kalacheva}}, \bibinfo {author} {\bibfnamefont {G.~P.}\ \bibnamefont {Fedorov}},\ and\ \bibinfo {author} {\bibfnamefont {O.~V.}\ \bibnamefont {Astafiev}},\ }\bibfield  {title} {\bibinfo {title} {Overcoming photon blockade in a circuit-qed single-atom maser with engineered metastability and strong coupling},\ }\href {https://doi.org/10.1103/PhysRevA.107.L031701} {\bibfield  {journal} {\bibinfo  {journal} {Phys. Rev. A}\ }\textbf {\bibinfo {volume} {107}},\ \bibinfo {pages} {L031701} (\bibinfo {year} {2023})}\BibitemShut {NoStop}%
\bibitem [{\citenamefont {Jaseem}\ \emph {et~al.}(2020{\natexlab{b}})\citenamefont {Jaseem}, \citenamefont {Hajdu\ifmmode~\check{s}\else \v{s}\fi{}ek}, \citenamefont {Vedral}, \citenamefont {Fazio}, \citenamefont {Kwek},\ and\ \citenamefont {Vinjanampathy}}]{jaseem2020quantum}%
  \BibitemOpen
  \bibfield  {author} {\bibinfo {author} {\bibfnamefont {N.}~\bibnamefont {Jaseem}}, \bibinfo {author} {\bibfnamefont {M.}~\bibnamefont {Hajdu\ifmmode~\check{s}\else \v{s}\fi{}ek}}, \bibinfo {author} {\bibfnamefont {V.}~\bibnamefont {Vedral}}, \bibinfo {author} {\bibfnamefont {R.}~\bibnamefont {Fazio}}, \bibinfo {author} {\bibfnamefont {L.-C.}\ \bibnamefont {Kwek}},\ and\ \bibinfo {author} {\bibfnamefont {S.}~\bibnamefont {Vinjanampathy}},\ }\bibfield  {title} {\bibinfo {title} {Quantum synchronization in nanoscale heat engines},\ }\href {https://doi.org/10.1103/PhysRevE.101.020201} {\bibfield  {journal} {\bibinfo  {journal} {Phys. Rev. E}\ }\textbf {\bibinfo {volume} {101}},\ \bibinfo {pages} {020201} (\bibinfo {year} {2020}{\natexlab{b}})}\BibitemShut {NoStop}%
\bibitem [{\citenamefont {Murtadho}\ \emph {et~al.}(2023{\natexlab{b}})\citenamefont {Murtadho}, \citenamefont {Thingna},\ and\ \citenamefont {Vinjanampathy}}]{murtadho2023PRA}%
  \BibitemOpen
  \bibfield  {author} {\bibinfo {author} {\bibfnamefont {T.}~\bibnamefont {Murtadho}}, \bibinfo {author} {\bibfnamefont {J.}~\bibnamefont {Thingna}},\ and\ \bibinfo {author} {\bibfnamefont {S.}~\bibnamefont {Vinjanampathy}},\ }\bibfield  {title} {\bibinfo {title} {Deriving lower bounds on the efficiency of near-degenerate thermal machines via synchronization},\ }\href {https://doi.org/10.1103/PhysRevA.108.012205} {\bibfield  {journal} {\bibinfo  {journal} {Phys. Rev. A}\ }\textbf {\bibinfo {volume} {108}},\ \bibinfo {pages} {012205} (\bibinfo {year} {2023}{\natexlab{b}})}\BibitemShut {NoStop}%
\bibitem [{\citenamefont {Lee~Loh}\ and\ \citenamefont {Kim}(2015)}]{lee2015visualizing}%
  \BibitemOpen
  \bibfield  {author} {\bibinfo {author} {\bibfnamefont {Y.}~\bibnamefont {Lee~Loh}}\ and\ \bibinfo {author} {\bibfnamefont {M.}~\bibnamefont {Kim}},\ }\bibfield  {title} {\bibinfo {title} {Visualizing spin states using the spin coherent state representation},\ }\href@noop {} {\bibfield  {journal} {\bibinfo  {journal} {American Journal of Physics}\ }\textbf {\bibinfo {volume} {83}},\ \bibinfo {pages} {30} (\bibinfo {year} {2015})}\BibitemShut {NoStop}%
\bibitem [{\citenamefont {Li}\ \emph {et~al.}(2020)\citenamefont {Li}, \citenamefont {Cai}, \citenamefont {Yan}, \citenamefont {Wang}, \citenamefont {Pan}, \citenamefont {Ma}, \citenamefont {Cai}, \citenamefont {Han}, \citenamefont {Hua}, \citenamefont {Han} \emph {et~al.}}]{li2020tunable}%
  \BibitemOpen
  \bibfield  {author} {\bibinfo {author} {\bibfnamefont {X.}~\bibnamefont {Li}}, \bibinfo {author} {\bibfnamefont {T.}~\bibnamefont {Cai}}, \bibinfo {author} {\bibfnamefont {H.}~\bibnamefont {Yan}}, \bibinfo {author} {\bibfnamefont {Z.}~\bibnamefont {Wang}}, \bibinfo {author} {\bibfnamefont {X.}~\bibnamefont {Pan}}, \bibinfo {author} {\bibfnamefont {Y.}~\bibnamefont {Ma}}, \bibinfo {author} {\bibfnamefont {W.}~\bibnamefont {Cai}}, \bibinfo {author} {\bibfnamefont {J.}~\bibnamefont {Han}}, \bibinfo {author} {\bibfnamefont {Z.}~\bibnamefont {Hua}}, \bibinfo {author} {\bibfnamefont {X.}~\bibnamefont {Han}}, \emph {et~al.},\ }\bibfield  {title} {\bibinfo {title} {Tunable coupler for realizing a controlled-phase gate with dynamically decoupled regime in a superconducting circuit},\ }\href@noop {} {\bibfield  {journal} {\bibinfo  {journal} {Physical Review Applied}\ }\textbf {\bibinfo {volume} {14}},\ \bibinfo {pages} {024070} (\bibinfo {year} {2020})}\BibitemShut {NoStop}%
\bibitem [{\citenamefont {Yan}\ \emph {et~al.}(2018)\citenamefont {Yan}, \citenamefont {Krantz}, \citenamefont {Sung}, \citenamefont {Kjaergaard}, \citenamefont {Campbell}, \citenamefont {Orlando}, \citenamefont {Gustavsson},\ and\ \citenamefont {Oliver}}]{yan2018tunable}%
  \BibitemOpen
  \bibfield  {author} {\bibinfo {author} {\bibfnamefont {F.}~\bibnamefont {Yan}}, \bibinfo {author} {\bibfnamefont {P.}~\bibnamefont {Krantz}}, \bibinfo {author} {\bibfnamefont {Y.}~\bibnamefont {Sung}}, \bibinfo {author} {\bibfnamefont {M.}~\bibnamefont {Kjaergaard}}, \bibinfo {author} {\bibfnamefont {D.~L.}\ \bibnamefont {Campbell}}, \bibinfo {author} {\bibfnamefont {T.~P.}\ \bibnamefont {Orlando}}, \bibinfo {author} {\bibfnamefont {S.}~\bibnamefont {Gustavsson}},\ and\ \bibinfo {author} {\bibfnamefont {W.~D.}\ \bibnamefont {Oliver}},\ }\bibfield  {title} {\bibinfo {title} {Tunable coupling scheme for implementing high-fidelity two-qubit gates},\ }\href@noop {} {\bibfield  {journal} {\bibinfo  {journal} {Physical Review Applied}\ }\textbf {\bibinfo {volume} {10}},\ \bibinfo {pages} {054062} (\bibinfo {year} {2018})}\BibitemShut {NoStop}%
\bibitem [{\citenamefont {Sung}\ \emph {et~al.}(2021)\citenamefont {Sung}, \citenamefont {Ding}, \citenamefont {Braum\"uller}, \citenamefont {Veps\"al\"ainen}, \citenamefont {Kannan}, \citenamefont {Kjaergaard}, \citenamefont {Greene}, \citenamefont {Samach}, \citenamefont {McNally}, \citenamefont {Kim}, \citenamefont {Melville}, \citenamefont {Niedzielski}, \citenamefont {Schwartz}, \citenamefont {Yoder}, \citenamefont {Orlando}, \citenamefont {Gustavsson},\ and\ \citenamefont {Oliver}}]{PhysRevX.11.021058}%
  \BibitemOpen
  \bibfield  {author} {\bibinfo {author} {\bibfnamefont {Y.}~\bibnamefont {Sung}}, \bibinfo {author} {\bibfnamefont {L.}~\bibnamefont {Ding}}, \bibinfo {author} {\bibfnamefont {J.}~\bibnamefont {Braum\"uller}}, \bibinfo {author} {\bibfnamefont {A.}~\bibnamefont {Veps\"al\"ainen}}, \bibinfo {author} {\bibfnamefont {B.}~\bibnamefont {Kannan}}, \bibinfo {author} {\bibfnamefont {M.}~\bibnamefont {Kjaergaard}}, \bibinfo {author} {\bibfnamefont {A.}~\bibnamefont {Greene}}, \bibinfo {author} {\bibfnamefont {G.~O.}\ \bibnamefont {Samach}}, \bibinfo {author} {\bibfnamefont {C.}~\bibnamefont {McNally}}, \bibinfo {author} {\bibfnamefont {D.}~\bibnamefont {Kim}}, \bibinfo {author} {\bibfnamefont {A.}~\bibnamefont {Melville}}, \bibinfo {author} {\bibfnamefont {B.~M.}\ \bibnamefont {Niedzielski}}, \bibinfo {author} {\bibfnamefont {M.~E.}\ \bibnamefont {Schwartz}}, \bibinfo {author} {\bibfnamefont {J.~L.}\ \bibnamefont {Yoder}}, \bibinfo {author} {\bibfnamefont {T.~P.}\ \bibnamefont {Orlando}}, \bibinfo {author}
  {\bibfnamefont {S.}~\bibnamefont {Gustavsson}},\ and\ \bibinfo {author} {\bibfnamefont {W.~D.}\ \bibnamefont {Oliver}},\ }\bibfield  {title} {\bibinfo {title} {Realization of high-fidelity cz and $zz$-free iswap gates with a tunable coupler},\ }\href {https://doi.org/10.1103/PhysRevX.11.021058} {\bibfield  {journal} {\bibinfo  {journal} {Phys. Rev. X}\ }\textbf {\bibinfo {volume} {11}},\ \bibinfo {pages} {021058} (\bibinfo {year} {2021})}\BibitemShut {NoStop}%
\bibitem [{\citenamefont {Li}\ \emph {et~al.}(2017)\citenamefont {Li}, \citenamefont {Xue}, \citenamefont {Tan}, \citenamefont {Liu}, \citenamefont {Dai}, \citenamefont {Zhang}, \citenamefont {Yu},\ and\ \citenamefont {Yu}}]{li2017APL}%
  \BibitemOpen
  \bibfield  {author} {\bibinfo {author} {\bibfnamefont {M.}~\bibnamefont {Li}}, \bibinfo {author} {\bibfnamefont {G.}~\bibnamefont {Xue}}, \bibinfo {author} {\bibfnamefont {X.}~\bibnamefont {Tan}}, \bibinfo {author} {\bibfnamefont {Q.}~\bibnamefont {Liu}}, \bibinfo {author} {\bibfnamefont {K.}~\bibnamefont {Dai}}, \bibinfo {author} {\bibfnamefont {K.}~\bibnamefont {Zhang}}, \bibinfo {author} {\bibfnamefont {H.}~\bibnamefont {Yu}},\ and\ \bibinfo {author} {\bibfnamefont {Y.}~\bibnamefont {Yu}},\ }\bibfield  {title} {\bibinfo {title} {{Two-qubit state tomography with ensemble average in coupled superconducting qubits}},\ }\bibfield  {journal} {\bibinfo  {journal} {Applied Physics Letters}\ }\textbf {\bibinfo {volume} {110}},\ \href {https://doi.org/10.1063/1.4979652} {10.1063/1.4979652} (\bibinfo {year} {2017}),\ \bibinfo {note} {132602}\BibitemShut {NoStop}%
\bibitem [{\citenamefont {Bohnet}\ \emph {et~al.}(2012)\citenamefont {Bohnet}, \citenamefont {Chen}, \citenamefont {Weiner}, \citenamefont {Meiser}, \citenamefont {Holland},\ and\ \citenamefont {Thompson}}]{bohnet2012Nat}%
  \BibitemOpen
  \bibfield  {author} {\bibinfo {author} {\bibfnamefont {J.~G.}\ \bibnamefont {Bohnet}}, \bibinfo {author} {\bibfnamefont {Z.}~\bibnamefont {Chen}}, \bibinfo {author} {\bibfnamefont {J.~M.}\ \bibnamefont {Weiner}}, \bibinfo {author} {\bibfnamefont {D.}~\bibnamefont {Meiser}}, \bibinfo {author} {\bibfnamefont {M.~J.}\ \bibnamefont {Holland}},\ and\ \bibinfo {author} {\bibfnamefont {J.~K.}\ \bibnamefont {Thompson}},\ }\bibfield  {title} {\bibinfo {title} {A steady-state superradiant laser with less than one intracavity photon},\ }\href {https://doi.org/10.1038/nature10920} {\bibfield  {journal} {\bibinfo  {journal} {Nature}\ }\textbf {\bibinfo {volume} {484}},\ \bibinfo {pages} {78} (\bibinfo {year} {2012})}\BibitemShut {NoStop}%
\bibitem [{\citenamefont {Xu}\ \emph {et~al.}(2014)\citenamefont {Xu}, \citenamefont {Tieri}, \citenamefont {Fine}, \citenamefont {Thompson},\ and\ \citenamefont {Holland}}]{xu2014synchronization}%
  \BibitemOpen
  \bibfield  {author} {\bibinfo {author} {\bibfnamefont {M.}~\bibnamefont {Xu}}, \bibinfo {author} {\bibfnamefont {D.~A.}\ \bibnamefont {Tieri}}, \bibinfo {author} {\bibfnamefont {E.}~\bibnamefont {Fine}}, \bibinfo {author} {\bibfnamefont {J.~K.}\ \bibnamefont {Thompson}},\ and\ \bibinfo {author} {\bibfnamefont {M.~J.}\ \bibnamefont {Holland}},\ }\bibfield  {title} {\bibinfo {title} {Synchronization of two ensembles of atoms},\ }\href@noop {} {\bibfield  {journal} {\bibinfo  {journal} {Physical review letters}\ }\textbf {\bibinfo {volume} {113}},\ \bibinfo {pages} {154101} (\bibinfo {year} {2014})}\BibitemShut {NoStop}%
\bibitem [{\citenamefont {Weiner}\ \emph {et~al.}(2017)\citenamefont {Weiner}, \citenamefont {Cox}, \citenamefont {Bohnet},\ and\ \citenamefont {Thompson}}]{weiner2017PRA}%
  \BibitemOpen
  \bibfield  {author} {\bibinfo {author} {\bibfnamefont {J.~M.}\ \bibnamefont {Weiner}}, \bibinfo {author} {\bibfnamefont {K.~C.}\ \bibnamefont {Cox}}, \bibinfo {author} {\bibfnamefont {J.~G.}\ \bibnamefont {Bohnet}},\ and\ \bibinfo {author} {\bibfnamefont {J.~K.}\ \bibnamefont {Thompson}},\ }\bibfield  {title} {\bibinfo {title} {Phase synchronization inside a superradiant laser},\ }\href {https://doi.org/10.1103/PhysRevA.95.033808} {\bibfield  {journal} {\bibinfo  {journal} {Phys. Rev. A}\ }\textbf {\bibinfo {volume} {95}},\ \bibinfo {pages} {033808} (\bibinfo {year} {2017})}\BibitemShut {NoStop}%
\bibitem [{\citenamefont {Zhu}\ \emph {et~al.}(2015)\citenamefont {Zhu}, \citenamefont {Schachenmayer}, \citenamefont {Xu}, \citenamefont {Herrera}, \citenamefont {Restrepo}, \citenamefont {Holland},\ and\ \citenamefont {Rey}}]{zhu2015NJP}%
  \BibitemOpen
  \bibfield  {author} {\bibinfo {author} {\bibfnamefont {B.}~\bibnamefont {Zhu}}, \bibinfo {author} {\bibfnamefont {J.}~\bibnamefont {Schachenmayer}}, \bibinfo {author} {\bibfnamefont {M.}~\bibnamefont {Xu}}, \bibinfo {author} {\bibfnamefont {F.}~\bibnamefont {Herrera}}, \bibinfo {author} {\bibfnamefont {J.~G.}\ \bibnamefont {Restrepo}}, \bibinfo {author} {\bibfnamefont {M.~J.}\ \bibnamefont {Holland}},\ and\ \bibinfo {author} {\bibfnamefont {A.~M.}\ \bibnamefont {Rey}},\ }\bibfield  {title} {\bibinfo {title} {Synchronization of interacting quantum dipoles},\ }\href {https://doi.org/10.1088/1367-2630/17/8/083063} {\bibfield  {journal} {\bibinfo  {journal} {New Journal of Physics}\ }\textbf {\bibinfo {volume} {17}},\ \bibinfo {pages} {083063} (\bibinfo {year} {2015})}\BibitemShut {NoStop}%
\bibitem [{\citenamefont {Shankar}\ \emph {et~al.}(2017)\citenamefont {Shankar}, \citenamefont {Cooper}, \citenamefont {Bohnet}, \citenamefont {Bollinger},\ and\ \citenamefont {Holland}}]{shankar2017PRA}%
  \BibitemOpen
  \bibfield  {author} {\bibinfo {author} {\bibfnamefont {A.}~\bibnamefont {Shankar}}, \bibinfo {author} {\bibfnamefont {J.}~\bibnamefont {Cooper}}, \bibinfo {author} {\bibfnamefont {J.~G.}\ \bibnamefont {Bohnet}}, \bibinfo {author} {\bibfnamefont {J.~J.}\ \bibnamefont {Bollinger}},\ and\ \bibinfo {author} {\bibfnamefont {M.}~\bibnamefont {Holland}},\ }\bibfield  {title} {\bibinfo {title} {Steady-state spin synchronization through the collective motion of trapped ions},\ }\href {https://doi.org/10.1103/PhysRevA.95.033423} {\bibfield  {journal} {\bibinfo  {journal} {Phys. Rev. A}\ }\textbf {\bibinfo {volume} {95}},\ \bibinfo {pages} {033423} (\bibinfo {year} {2017})}\BibitemShut {NoStop}%
\bibitem [{\citenamefont {Landa}\ and\ \citenamefont {Misguich}(2023)}]{landa2023SciPost}%
  \BibitemOpen
  \bibfield  {author} {\bibinfo {author} {\bibfnamefont {H.}~\bibnamefont {Landa}}\ and\ \bibinfo {author} {\bibfnamefont {G.}~\bibnamefont {Misguich}},\ }\bibfield  {title} {\bibinfo {title} {{Nonlocal correlations in noisy multiqubit systems simulated using matrix product operators}},\ }\href {https://doi.org/10.21468/SciPostPhysCore.6.2.037} {\bibfield  {journal} {\bibinfo  {journal} {SciPost Phys. Core}\ }\textbf {\bibinfo {volume} {6}},\ \bibinfo {pages} {037} (\bibinfo {year} {2023})}\BibitemShut {NoStop}%
\bibitem [{\citenamefont {Bera}\ \emph {et~al.}(2023)\citenamefont {Bera}, \citenamefont {Pandit}, \citenamefont {Chatterjee}, \citenamefont {Singh}, \citenamefont {Lewenstein}, \citenamefont {Bhattacharya},\ and\ \citenamefont {Bera}}]{bera2023arXiv}%
  \BibitemOpen
  \bibfield  {author} {\bibinfo {author} {\bibfnamefont {M.~L.}\ \bibnamefont {Bera}}, \bibinfo {author} {\bibfnamefont {T.}~\bibnamefont {Pandit}}, \bibinfo {author} {\bibfnamefont {K.}~\bibnamefont {Chatterjee}}, \bibinfo {author} {\bibfnamefont {V.}~\bibnamefont {Singh}}, \bibinfo {author} {\bibfnamefont {M.}~\bibnamefont {Lewenstein}}, \bibinfo {author} {\bibfnamefont {U.}~\bibnamefont {Bhattacharya}},\ and\ \bibinfo {author} {\bibfnamefont {M.~N.}\ \bibnamefont {Bera}},\ }\href@noop {} {\bibinfo {title} {Steady-state quantum thermodynamics with synthetic negative temperatures}} (\bibinfo {year} {2023}),\ \Eprint {https://arxiv.org/abs/2305.01215} {arXiv:2305.01215 [quant-ph]} \BibitemShut {NoStop}%
\bibitem [{\citenamefont {Johansson}\ \emph {et~al.}(2012)\citenamefont {Johansson}, \citenamefont {Nation},\ and\ \citenamefont {Nori}}]{johansson2012qutip}%
  \BibitemOpen
  \bibfield  {author} {\bibinfo {author} {\bibfnamefont {J.~R.}\ \bibnamefont {Johansson}}, \bibinfo {author} {\bibfnamefont {P.~D.}\ \bibnamefont {Nation}},\ and\ \bibinfo {author} {\bibfnamefont {F.}~\bibnamefont {Nori}},\ }\bibfield  {title} {\bibinfo {title} {{{QuTiP}}: {{An}} open-source {{Python}} framework for the dynamics of open quantum systems},\ }\href@noop {} {\bibfield  {journal} {\bibinfo  {journal} {Computer Physics Communications}\ }\textbf {\bibinfo {volume} {183}},\ \bibinfo {pages} {1760} (\bibinfo {year} {2012})}\BibitemShut {NoStop}%
\bibitem [{\citenamefont {Dowling}\ \emph {et~al.}(1994)\citenamefont {Dowling}, \citenamefont {Agarwal},\ and\ \citenamefont {Schleich}}]{dowling1994wigner}%
  \BibitemOpen
  \bibfield  {author} {\bibinfo {author} {\bibfnamefont {J.~P.}\ \bibnamefont {Dowling}}, \bibinfo {author} {\bibfnamefont {G.~S.}\ \bibnamefont {Agarwal}},\ and\ \bibinfo {author} {\bibfnamefont {W.~P.}\ \bibnamefont {Schleich}},\ }\bibfield  {title} {\bibinfo {title} {Wigner distribution of a general angular-momentum state: Applications to a collection of two-level atoms},\ }\href@noop {} {\bibfield  {journal} {\bibinfo  {journal} {Physical Review A}\ }\textbf {\bibinfo {volume} {49}},\ \bibinfo {pages} {4101} (\bibinfo {year} {1994})}\BibitemShut {NoStop}%
\bibitem [{\citenamefont {Agarwal}(1981)}]{agarwal1981relation}%
  \BibitemOpen
  \bibfield  {author} {\bibinfo {author} {\bibfnamefont {G.~S.}\ \bibnamefont {Agarwal}},\ }\bibfield  {title} {\bibinfo {title} {Relation between atomic coherent-state representation, state multipoles, and generalized phase-space distributions},\ }\href@noop {} {\bibfield  {journal} {\bibinfo  {journal} {Physical Review A}\ }\textbf {\bibinfo {volume} {24}},\ \bibinfo {pages} {2889} (\bibinfo {year} {1981})}\BibitemShut {NoStop}%
\end{thebibliography}

\providecommand{\noopsort}[1]{}\providecommand{\singleletter}[1]{#1}%

\end{document}